\documentclass[twocolumn,showpacs,preprintnumbers,amsmath,amssymb]{revtex4}

	\newif\ifpdf
	\ifx\pdfoutput\undefined
	\pdffalse 
	\else
	\pdfoutput=1 
	\pdftrue
	\fi

	\ifpdf
	\usepackage[pdftex]{graphicx}
	\else
	\usepackage{graphicx}
	\fi

	\ifpdf
	\DeclareGraphicsExtensions{.pdf, .jpg}
	\else
	\DeclareGraphicsExtensions{.eps, .jpg}
	\fi

\usepackage{graphicx}
\usepackage{dcolumn}
\usepackage{bm}

\begin{document}
\preprint{submitted to Phys. Rev. E}

\def\r{{\bf r}}

\title{Effects of {\it inhomogeneous} partial absorption and the geometry of the boundary on the population evolution of molecules diffusing in general porous media}

\author{Seungoh Ryu}
\email{sryu@slb.com}
\affiliation{%
Schlumberger Doll Research\\
One Hampshire Street, Cambridge, MA 02139, USA
}%
\date{\today}
\begin{abstract}
We consider aspects of the population dynamics, inside a bound domain, of diffusing agents carrying an attribute which is stochastically destroyed upon contact with the boundary. 
The normal mode analysis of the relevant Helmholtz equation under the partially absorbing, but uniform, boundary condition provides a starting framework in understanding detailed evolution dynamics of the attribute in the time domain. In particular, the boundary-localized depletion has been widely employed in practical applications that depend on geometry of various porous media such as rocks, cement, bones, and cheese. 
While direct relationship between the pore geometry and the diffusion-relaxation spectrum forms the basis for such applications and has been extensively studied, relatively less attention has been paid to the spatial variation of the boundary condition. 
In this work, we focus on the way the pore geometry and the {\em inhomogeneous} depletion strength of the boundary become intertwined and thus obscure the direct relationship between the spectrum and the geometry. It is often impossible to gauge experimentally the degree to which such interference occur.  
We fill this gap by perturbatively incorporating classes of spatially-varying boundary conditions and derive their consequences that are observable through numerical simulations or controlled experiments on glass bead packs and artificially fabricated porous media. 
We identify features of the spectrum that are most sensitive to the inhomogeneity and apply the method  to the spherical pore with a simple hemi-spherical binary distribution of the depletion strength and obtain bounds for the induced change in the slowest relaxation mode. 
\end{abstract}


\maketitle
\section{\label{sec:introduction}Introduction}
We consider the evolution of a physical attribute carried by  a population of random-walkers inside a medium bound by a wall of general shape. When a walker hits the boundary, the encounter depletes its attribute with a certain probability, $p \in [0, 1]$. Our main concern here is on allowing this probability to have general spatial variation and to investigate its consequences on the spatio-temporal evolution of the local attribute density $\Psi(\r, t)$ and its net sum, ${\cal M} (t) \equiv \int  d\r \Psi(\r, t)$. 
Without the spatial variations of $p$ and the local diffusivity, $D$, the problem reduces to the classic Helmholtz equation with a {\it uniform} Robin's boundary condition, bookended by the Dirchlet- ($p  \rightarrow \infty$) on one limit and by the Neumann- condition ($p \rightarrow 0$) on the other. The spectral analysis of its eigenmodes has been discussed as a probe of geometrical properties of the boundary\cite{Kac1966,Gordon1992,Chapman1995} and found application for a variety of systems such as the electrode impedance\cite{Sapoval1999}, acoustics\cite{Rocchesso2001}, NMR relaxometry\cite{deGennes1982,Mitra1992}, nuclear level statistics\cite{Metha2004}, quantum chaos\cite{da1997}, and migration of cultural or genetic trait\cite{Fisher1937,Fort1999a,Fort1999}. Population evolution of the Web crawler-programs\cite{Takeno2006} deployed over a large network in the presence of unstable nodes may be an example where the diffusion may not necessarily be bound to the physical space.

To be concrete, we are directly motivated by issues encountered in the interpretation of the magnetic resonance (MR) probe of fluid in conventional porous media\cite{Grebenkov2007,Kleinberg1996}, suspended particulate aggregates and colloids\cite{Bloembergen1948, Perez2002, Kansal2002}.  
The diffusion-relaxation dynamics of polarized proton spins carried by diffusing molecules has been widely exploited, sometimes without full justification, to characterize the pore geometry and fluid viscosity in the soil\cite{Bryar2002,Bryar2008}, cements\cite{Blinc1988}, oil pigment of old paintings\cite{Casieri2005}, biological tissues\cite{Fenrich2001}, fiber bundles\cite{Bronskill1994}, plant cells\cite{Weerd2002a} or a piece of cheese\cite{Hurlimann2006}. 
In its geophysical or oil-field application, the interface-enhanced relaxation is used as a probe for the pore geometry of rocks\cite{Kenyon1992,Kleinberg1996,Sorland2007}, composition of pore filling fluids\cite{Hurlimann2002}, and even wetting conditions. This extraordinary utility derives from the basic observation that the interface-enhanced relaxation rate (widely called $T_2$ distribution \cite{Kleinberg1996}) is directly proportional to the surface-to-volume ratio of the pore enclosure when certain conditions are met. (See Eqs.\ref{eq:definekappa}-\ref{eq:tsv} below)

Three basic conditions are required to be met implicitly in such a mapping between the relaxation spectrum and the pore-size distribution: First, the porous medium is pictured as an aggregation of isolated pores, which allows an unambiguous notion of the pore size if individual pores are of simple geometry. This may not necessarily require each pore to be closed. A periodic, symmetric arrangement of pore space\cite{Bergman1995a} connected by narrow channels may well be considered as such, as long as the inter-pore diffusive flux either balances out or becomes negligible. Attempts to map between the pore size- and the surface-enhanced $T_2-$ distributions may work very well for systems such as mono-disperse bead packs, food elements composed of suspended spherical voids such as cheese, and  the class of sedimentary rocks such as  clean sandstones. 

Second, in the absence of diffusive flux among such pores, the so-called {\it fast-diffusion} condition is met so that the relaxation spectrum is dominated by the slowest mode for each pore the rate for which becomes proportional to the respective surface-to-volume ratio.  For a simple isolated spherical pore of radius $a$, for example, the condition involves a single dimensionless parameter\cite{Brownstein1979}
\begin{equation}
\label{eq:definekappa}
\kappa \equiv  \rho_0 a / D \ll 1.
\end{equation}
The macroscopic parameter $\rho_0$ characterizes the uniform depletion rate at the interface and is directly related to the probability of depletion $p$ and enter the Robin's condition in the form of 
$[D\hat{n} \cdot \nabla +\rho_0 ]\Psi(\r) = 0$ on the boundary with $\hat{n}$ being the unit vector normal to the interface. When this condition prevails, it was noted \cite{Brownstein1979} that 
the surface-induced depletion rate $w_s$ (or so-called surface-enhanced $T_2$ relaxation rate \cite{Kleinberg1996} in the NMR context) is directly related to the surface (S)-to-volume(${\cal V}$) ratio of the pore: 
\begin{equation}
\label{eq:tsv}
w_s = \rho_0 \frac{\cal S} {{\cal V}}.
\end{equation}
This simple relationship had been applied widely in examples mentioned above\cite{Bryar2002,Bryar2008, Blinc1988, Casieri2005, Fenrich2001, Bronskill1994, Weerd2002a, Hurlimann2006, Kenyon1992, Kleinberg1996, Sorland2007}.
For classes of porous media with a broad variation of its geometrical properties and strong diffusive coupling among its pore-constituents, these assumptions may break down. For example, in the oil-exploration, problems have been long recognized for the class of rocks in which the pore shape and the lithological composition of the matrix become complex. The complications induced by the heterogeneous, extended pore space in MR as well as other physical properties such as the electrical and hydraulic conductivity pose a fundamental challenge and invites active debates. 
As we will show in section \ref{sec:uniform}, this condition is facilitated by the {\it near-uniformity} of the slowest  eigenmode (See Eq.\ref{eq:excitedfraction}). When the pore geometry has more than one length scale (Eq.\ref{eq:generaldefinekappa}), the slowest mode acquires more pronounced spatial variation and the condition is relatively poorly met.

\begin{figure}
\includegraphics[width = 3.5in]{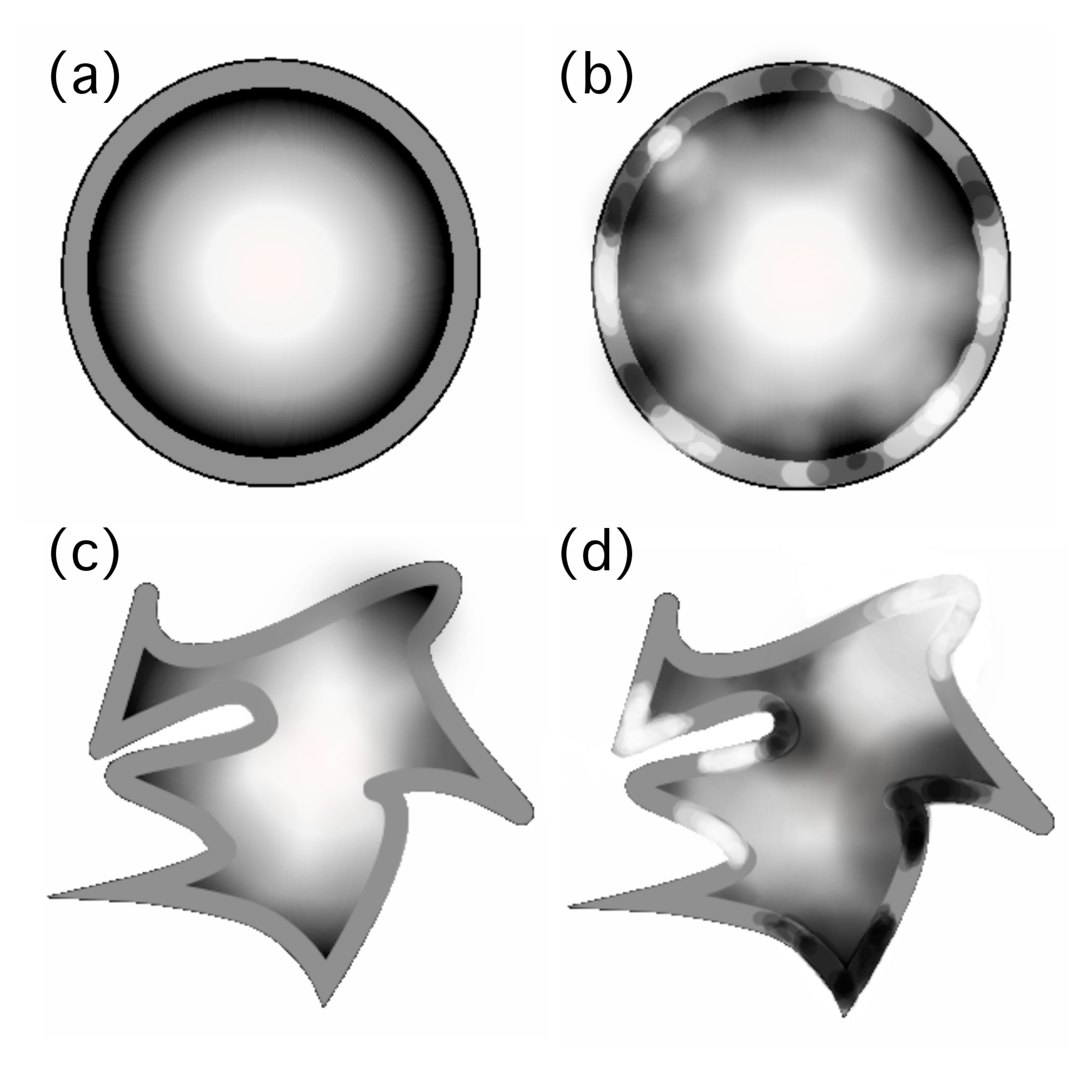}
\caption{\label{fig:cartoon} 
An artistic rendition of situations considered in the text. Panel (a) shows a simple spherical pore with a uniform $\rho_0$ (its strength indicated on the shell with a uniform gray color). Inside, the color represents the local population density (white:high, black:low) as evolved from a uniform initial distribution in the long time limit. Panel (b) shows the same, but with a non-uniform $\rho(\r)$ on the shell.  Panel (c) shows a more complex pore geometry but with uniform $\rho$. (d) shows the same with $\rho(\r)$ with a potential disruption on the registry between the local density and its pore-geometrical parameters. 
}
\end{figure}

The third assumption often made is that the surface-relaxation occurs with a uniform strength (i.e. $\rho(\r) = \rho_0$) throughout the interface even though an {\it inhomogeneous} $\rho$ is the norm, rather than an aberration,  in natural media.
In porous rocks, 
there are several mechanisms for the surface-enhanced relaxation\cite{Korb1996,DeSweet1994,Gillis1987,Dunn2002,Kleinberg1994}, and they often involve the stochastic distribution of magnetic minerals in the matrix or the local interfacial morphology. The strength of $\rho$ arising from such origins would acquire inhomogeneity across the pore-grain interface, but microscopic details of such a variation is not known quantitatively in general. 

Given the lack of such information, it is not entirely possible to dismiss the following observation: That for the population evolution of a collection of isolated pores of varying sizes, all satisfying the condition above (Eq.\ref{eq:definekappa}), it is possible to construct a collection of {\it identical} pores whose size $a_{min}$ is chosen such that $\rho_{max}  a_{min} / D \ll 1$, but with a distribution of $\rho$ values $\in [\rho_{min}, \rho_{max}]$ assigned to each, that will yield the identical evolution.  
It is worth noting that the question of whether there is a unique mapping between the eigenspectrum and a given boundary geometry has been posed in more abstract and {\it stronger} terms\cite{Kac1966,Gordon1992,Chapman1995}. 
The hypothetical situation for an NMR relaxometry as posed above violates a {\it weaker} form of {\it isospectral} criterion as it involves the behavior of the overall population decay (${\cal M}(t)$) obtained under both the uniform initial distribution and detection sensitivity profile (see Eq.\ref{eq;totalM} in the following) under the condition $\kappa \ll 1$.  
Aside from mathematical rigor, questions arise as to whether the direct mapping between the geometry and the relaxation rates could remain useful for a general $\rho(\r)$ profile. 
Figure \ref{fig:cartoon} summarizes the core issues in the form of the slowest mode profiles (casually rendered here for illustrative purpose) inside porous media with simple and moderately complex shapes (panels a and c). They are then further complicated by the presence of inhomogeneous $\rho(\r)$ which is {\it incommensurate} with the variation of the boundary shape (panels b and d).

Several authors had considered the effect of an inhomogeneous $\rho(\r)$.  Wilkinson {\it et al} incorporated the inhomogeneous $\rho$ in a toy model\cite{Wilkinson1991} in reduced dimensions. Kansal and Torquato considered a numerical technique to derive the {\it effective} trapping rate for a mixture of partially aborbing traps in the context of biological systems\cite{Kansal2002}. 
Valfouskaya  {\it et al}\cite{Valfouskaya2006}  considered a non-uniform absorption on randomly distributed sites in reconstructed porous media.  Arns {\it et al} used numerical simulations with a focus on the cross-correlation between the relaxation spectrum and the transport property\cite{Arns2006}. While the latter touches directly on one of the important practical issues, it provides little insight beyond the complications due to the pore-geometry issue (first and second conditions) alone.

\section{Setting up the problem}
\label{sec:defineproblem}
This work is concerned with consequences of allowing either of or all three assumptions above to break down. 
We aim to develop a method that incorporates the two components (geometry vs. inhomogeneous $\rho$) on an equal footing. 
To be precise, consider a pore space (${\cal V}$ denotes its pore-volume in the following) defined by the solid(grain)-pore interface $\Sigma$, its area designated as $S$) in a Euclidean space of dimension $d_e$. A physical property (such as polarized spin) is carried by molecules (or agents) diffusing through ${\cal V}$ with its mobility characterized by the local diffusion tensor $D(\r)$. We allow such molecules to get absorbed (or killed) by a certain mechanism at the boundary, if the property we are tracking is their population density, or allow the physical property to be drained upon contacting the interface with a certain probability. The strength of such surface-localized depletion mechanism is controlled by a parameter $\rho (\r)$, (See Eqs.\ref{eq:uniformbc} and \ref{eq:newbc}) which defines the coarse-grained strength of absorption/depletion/relaxation. 
For an isotropic system, the probability of depletion per collision with  the boundary, $p$, is related to $\rho$   via\cite{Wilkinson1991,Mendelson1993,Ryu2008b} 
$\rho (\r) = \sqrt{d_e} \frac{3}{2} \frac{D}{\epsilon} \frac{p}{1 - p/ 2}$
where the Brownian particle moves with continuous step sizes uniformly distributed in the interval $[-\epsilon, \epsilon]$ for each of the $d_e$ directions during the time step. Inhomogeneity in $\rho(\r)$ may arise through spatial variation in $p$ and/or $D/\epsilon$. 
The microscopic mechanism for the draining probability ($p$) will affect the {\em texture}, the spatial prifile, of $\rho(\r)$, but we will derive our main results without assuming a specific pattern for $\rho(\r)$.  Using the standard bra-ket notation\cite{Dirac1958,Arfken1970} the local population distribution at time $t$ is represented in terms of the basis functions $\{ | \r> \} $ as  $<\r |\Psi>_t (\equiv \Psi(\r, t))$, and overlap integral between two such functions $<\Psi | \Psi'>$ is equivalent to $\int_{{\cal V}} d\r \Psi^*(\r) \Psi(\r)$. The basis functions $\{ | \r> \} $ satisfy the orthogonal property:
$<\r' | \r > = \delta (\r - \r')$
where $ \delta (\r - \r')$ is the $d_e$-dimensional Dirac-delta function with the normalization $\int_{{\cal V}} \delta (\r - \r') d\r = 1$, integrated over the pore volume ${\cal V}$ and $\r' \in {\cal V}$. 
In the following section, we consider the diffusion equation according to which an initial state $| \Psi>_{t=0}$ evolves and consider the spectral property of the associated boundary value problem.  Specifics of pore shape variation is incorporated into more generic spectral features of the modes. The varying degree of break down for the second and the third conditions is then systematically studied via the changes reflected on the spectra for a range of values in $\kappa$ (Eqs.\ref{eq:definekappa} and \ref{eq:generaldefinekappa}) and the dimensionless parameter 
\begin{equation}\label{eq:definedeltarho}
\sigma = \frac{<|\delta \rho (\r)|>}{\rho_0} \equiv \frac{<| \rho(\r)-\rho_0|>}{\rho_0} ,
\end{equation}
with $\rho_0$ being the interfacial average of $\rho(\r)$. Obviously, $<\delta \rho(\r)> = 0$.
What are the main observable consequences for allowing $\sigma \ne 0$ in a natural porous media? 
In this work, we focus on the changes in the eigenvalue and the spatial mode profile of the slowest mode treating $\sigma$ as the small parameter.  We point out that this was largely motivated through mapping our Helmholtz problem to that of Schr\"odinger problem for the particle in a partially absorbing box in the imaginary time domain and treating $\delta \rho(\r)$ as a perturbative potential\cite{Ryu2001}. The mapping underscores the significance of the statistics and symmetries of the modes, especially the ground state, which are not as readily apparent in the traditional Green's function formalism.  
This then prompts us to ask whether the effects should be more pronounced where the inhomogeneity is commensurate with the variations in the underlying boundary shape, which in turn affects the spatial profile of the modes. As a corollary, it follows that  the faster modes, having little spectral overlap with the spatial variation of $\rho(\r)$, unless $\rho(\r)$ is self-affine, may be less sensitive compared to their slower counterparts. These aspects should be considered on an equal footing along with the diffusive coupling\cite{McCall1991,Song2000,Zielinski2002} in affecting the spectral properties of the problem as we will elaborate in sections \ref{sec:uniform} and \ref{sec:theory}. For a class of experimental diffusion probes, parallels with the spectroscopy of a quantum particle \cite{Ryu2001} yield useful insight and has led to novel applications\cite{Lisitza2002, Song2008} even with $\sigma=0$. 

The organization of the rest of the paper is as follows: We consider in section \ref{sec:uniform} the case of a uniform depletion strength $\rho_0$ on the boundary and offer an expanded account of observations (Eqs.\ref{eq:theorem},\ref{eq:defineell} and \ref{eq:excitedfraction}) that had been made earlier\cite{Ryu2001}) on the spectrum for a general boundary geometry with uniform $\rho$.  In section \ref{sec:theory}, we further develop for the spatially varying $\rho(\r)$, establishing a set of  fundamental relationships linking the uniform and non-uniform cases through a perturbative solution for the fractional changes of the eigenvalues and their weights. Eq.\ref{eq:fractionalshiftsmalldeltarho4} represents the central result of this perturbative approach. As a concrete example, we apply the method for  a spherical pore with a general angular variation of $\rho(\r)$ in section \ref{sec:sphere} and obtain solutions for a specific {\em binary} $\delta \rho(\r)$ texture. 

\section{\label{sec:uniform}Uniform $\rho$}
We start by considering the simpler case of a uniform $\rho (\r) = \rho_0$ and a general local diffusion tensor $D$.  
Introducing the flux operator ${\bf J}$ 
\begin{equation}
{\bf J} \equiv - D  \cdot \nabla 
\end{equation}
and the {\em Hamiltonian} operator ${\cal H}$:
\begin{equation}
\label{eq:hamiltonian}
{\cal H} \equiv \nabla \cdot  {\bf J}  =- \nabla \cdot D \cdot  \nabla ,
\end{equation}
the evolution of $\Psi$ as dictated by {\em continuity} follows
\begin{equation}
\label{eq:schrodinger}
\partial_t  | \Psi > =  - {\cal H} | \Psi >
\end{equation}
which formally links the slowest depletion rate of our diffusion-relaxation problem with the ground state energy of the analogous quantum mechanical system.\cite{Ryu2001} 
In real life MR relaxometry, diffusion of polarized spin-carrying molecules suffers an additional depletion in the bulk of the fluid if there exists a static field gradient. This {\it dephasing} may be eliminated via an experimental technique and therefore we neglect it for simplicity and consider only the depletion localized at the interface. 

Let us consider a partially absorbing boundary:
\begin{equation}
\label{eq:bc}
<\r | \hat{n} (\r) \cdot {\bf J} | \Psi >  =  \rho_0 < \r |  \Psi >  \quad {\rm for}\, \r \in \Sigma
\end{equation}
where $\hat{n}(\r)$ is the unit surface normal vector at the boundary point $\r$ pointing into the solid matrix. 
Time evolution of an initial distribution $|\Psi>_{0}$ can be expressed as a linear superposition of the set of eigenmodes $\{ | \phi_p^0 > \}$($p=0,1,2, \ldots$) of ${\cal H}$ 
\begin{equation}
|\Psi>_t = \sum_{p=0}^\infty | \phi_p^0> e^{-\lambda_p^0 t} <\phi_p^0 | \Psi>_{0}.
\end{equation}
Each eigenmode $|\phi_p^0>$ satisfies the equation
\begin{equation}
\label{eq:eigen}
{\cal H} |\phi_p^0> = \lambda_p^0 |\phi_p^0>
\end{equation}
and the boundary condition at interface $\Sigma$:
\begin{equation}
\label{eq:uniformbc}
<\r | \hat{n} (\r) \cdot {\bf J} | \phi_p^0 >  =  \rho_0 < \r  |\phi_p^0>  \quad {\rm for}\, \r \in \Sigma .
\end{equation}
Multiplying Eq.\ref{eq:eigen} by $<\phi_p^0|$, we obtain
\begin{equation}
< \phi_p^0  |  \nabla \cdot {\bf J}  | \phi_p^0>   = \lambda_p^0 
\end{equation}
where the left hand side, upon inserting the complete set of basis functions $I = | \r > \int d\r < \r |$, becomes the volume integration of 
\begin{equation}
\nabla \cdot (< \phi_p^0  | \r > < \r |   {\bf J} | \phi_p^0>) -  < \r |   {\bf J}  | \phi_p^0> \cdot  \nabla < \phi_p^0  | \r > .
\end{equation}
Combined with the boundary condition, we obtain the following expression for the eigenvalue: \cite{Ryu2001}
\begin{equation}
\label{eq:theorem}
\lambda_p^0 = << \phi_p^0  | \rho_0|  \phi_p^0 >> +  <\phi_p^0 | {\bf J} \cdot D^{-1} \cdot  {\bf J} | \phi_p^0>
\end{equation}
where $<<\ldots >> \equiv \oint_\Sigma \ldots d\sigma$. 
Employing the spatial representation, the right hand side is equivalent to
\begin{equation}
\label{eq:theoremalt}
\oint_\Sigma \rho_0 |\phi_p^0 (\r)|^2 d\sigma +  \int_{{\cal V}}  D_{\alpha\beta}(\r)  (\nabla_\alpha  \phi_p^0 ( \r )  ) (\nabla_\beta \phi_p^0 (\r) ) d\r 
\end{equation}
where $\alpha, \beta = x, y, z$. The result breaks the rate associated with each eigenmode into two channels:
a surface integral involving $\rho_0$ and a volume integral involving spatial variaion of the mode, ${\bf J} |\phi_p^0>$ and allows us to generalize the criterion of slow- and fast-diffusion regime\cite{Brownstein1979} beyond the simple pore geometry.  For the spherical pore, Brownstein and Tarr had shown that the dimensionless parameter $\kappa = \rho_0 a / D$ controls the qualitatively distinct behavior for the time evolution of $\Psi$. Based on the observation that the distinction originates from the degree of spatial fluctuation in the slowest mode (See Eq.\ref{eq:excitedfraction} below), we generalize the criterion by defining  $\kappa$ parameter as the ratio of the two terms in Eq.\ref{eq:theoremalt} for the slowest mode, $|\phi_0^0>$: 
\begin{equation}
\label{eq:generaldefinekappa}
\kappa \equiv   \frac{<\phi_0^0 | {\bf J} \cdot D^{-1} \cdot  {\bf J} | \phi_0^0>}{<< \phi_0^0  | \rho_0|  \phi_0^0 >>}
 = 
 \frac{\rho_0}{D} \frac{\int d\r (\nabla \phi_0^0)^2}{\oint d\sigma (\nabla \phi_0^0)^2}.
\end{equation}
If the diffusive flux in the bulk dominates, ($\kappa \rightarrow \infty$), the slowest rate becomes independent of $\rho_0$, while in the opposite limit, we have $\lambda_0^0 \rightarrow \rho_0 \oint d\sigma (\phi_0^0)^2 = \rho_0 S / {\cal V}.$
We therefore identify the length-scale parameter
\begin{equation}
\label{eq:defineell}
\ell \equiv 
\frac{\int d\r (\nabla \phi_0^0)^2} {\oint d\sigma (\nabla \phi_0^0)^2}, 
\end{equation}
for given $\rho_0$ and $D$, as the relevant {\em size} that separates the distinct regimes
for an arbitrary pore shape. It is interesting to note that $\ell$ is reminiscent of the $\Lambda$ parameter introduced by Johnson {\em et al}\cite{Johnson1986} in the context of electrical conductivity in general porous media.

Eq.\ref{eq:theorem} may be also used to investigate the effect of changes in $\rho_0$ and $D$ as induced via control parameters such as the temperature, $T$. We obtain
\begin{eqnarray}
&& \frac{d\lambda_p^0}{dT}  = << \phi_p^0  | \frac{d \rho_0(T)}{dT}  | \phi_p^0>>  + 2 << \phi_p^0  |  \rho_0  | \delta \phi_p^0 >> \\ 
&&+< \phi_p^0 |  \frac{d}{dT} {\bf J} \cdot D^{-1} \cdot  {\bf J}  |  \phi_p^0 > + 2 < \phi_p^0 |  {\bf J}\cdot D^{-1} \cdot  {\bf J} |  \delta \phi_p^0 > \nonumber 
\end{eqnarray}
where $|\delta \phi_p^0> = \frac{d}{dT}|\phi_p^0> $. Note that the slowest mode is the most sensitive to $\frac{d\rho_0}{dT}$ in the so-called {\it fast-diffusion} limit, where the contribution from the surface-integral dominates over the second term in Eq.\ref{eq:theoremalt}.

\begin{figure}
\includegraphics[width = 3.5in]{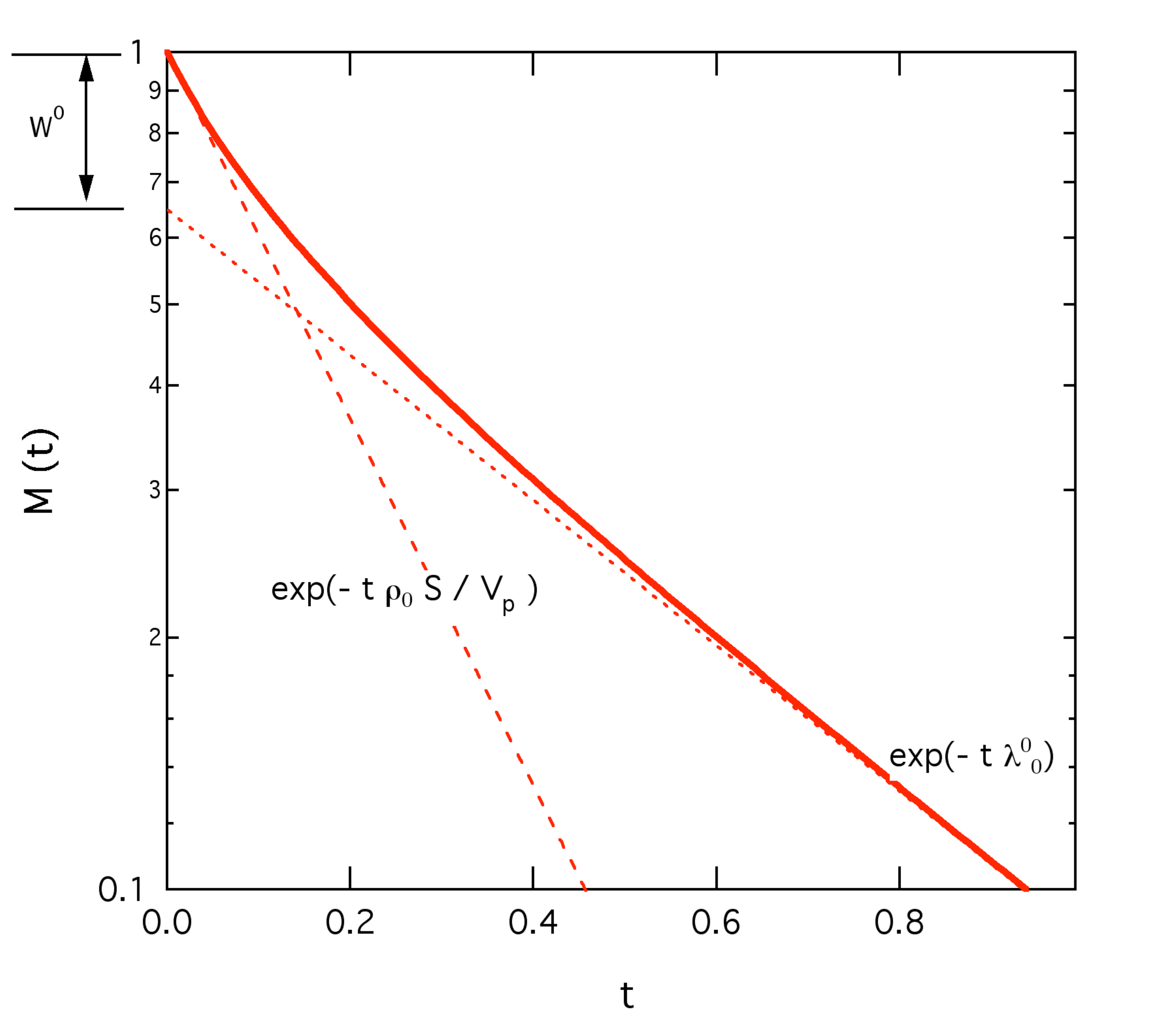}
\caption{\label{fig:schematics} A schematics of a typical evolution of a population $M$ with a uniform boundary condition with $\rho_0$. At early times, the slope matches that of $\rho_0 S / {\cal V}$, and later times, it approaches that of the slowest eigenmode with $\lambda_0^0$. (Each represented by broken curves). The spectral weight for the excited modes ($W^0$) is also indicated.}
\end{figure}

Here we also prove that  {\it the spectral weight of the excited modes in an initially uniform distribution is closely related to the spatial fluctuation of the slowest mode}.  To do so, let us take the initial state 
\begin{equation}
\label{eq:uniformstate}
\Psi_0 (\r) = \frac{1}{\sqrt{{\cal V}}} \Theta( \r \in {\cal V})
\end{equation}
where $\Theta(x) = 1$ if the boolean condition $x$ is satisfied, $0$ otherwise. 
Its subsequent time evolution follows 
\begin{equation}
\label{eq:Mt}
| \Psi>_t = \sum_q^{all} s_q^0 e^{-t  \lambda_q^0} |\phi_q^0>
\end{equation}
with its spectral distribution $s^0_q$ given by 
\begin{equation}
\label{eq:spectraldistribution}
s_q^0 = <\phi_q^0 | \Psi>_0 = \frac{1}{\sqrt{{\cal V}}} \int_{{\cal V}} \phi_q^0 (\r) d\r.
\end{equation}
The total population follows
\begin{equation}
\label{eq;totalM}
{\cal M}(t) =\frac{1}{\sqrt{{\cal V}}}  \int_{{\cal V}} \Psi_t(\r) d\r =  \sum_q^{all} |s_q^0|^2 e^{-t \lambda_q^0}.
\end{equation}
It can be shown that the fraction of the initial population belonging to the {\it excited} modes (i.e. $q \ne 0$) is 
\begin{equation}
\label{eq:excitedfraction}
W^0  \equiv  \sum_{q\ne 0} |s_q^0|^2 
= {\cal V} ( < |\phi_0^0|^2> - |<\phi_0^0>|^2 )
\end{equation}
which means that the total weight for all excited modes is directly proportional to the mean-square variance of the lowest mode. 
It is straightforward to show using Eqs. \ref{eq:schrodinger} and \ref{eq:bc},  that the slope at very early times, if the initial distribution is uniform, 
\begin{equation}
\label{eq:initialslope0}
- \frac{1}{{\cal M}(t)} \frac{d}{dt} {\cal M}(t) =   \frac{\int d\r  <\r |  {\cal H} | \Psi>_t}{\int d\r  <\r | \Psi>_t} 
=  \frac{\oint d\sigma \rho_0  <\r | \Psi>_t}{\int d\r  <\r | \Psi>_t} 
\end{equation}
which reduces to $ \rho_0 \frac{S}{{\cal V}}$
using the property that ${\cal M}(t)$ is uniform in the limit $t \rightarrow 0$. Note that the result applies also to the case of {\it disjoint} pore systems if one understands $S$ and ${\cal V}$ to represent the total interface area and the pore volume respectively. However, the range in time during which this assumption remains valid depends on the pore shape and is not a universal property as it depends on the boundary geometry of a given system. The following sum rule also follows:
\begin{equation}
\label{eq:sumrule0}
\sum_q^{all} |s_q^0|^2 \lambda_q^0 = \rho_0 \frac{S}{{\cal V}}
\end{equation}
and from the positive definiteness of $\lambda_q^0$'s, it also follows that  $\lambda_0^0$ is bounded
\begin{equation}
\label{eq:boundforlambda0}
\rho_0 \frac{S}{{\cal V}} \ge  \lambda_0^0.
\end{equation}
The properties described in this section are schematically depicted in Figure \ref{fig:schematics}.

\section{\label{sec:theory}Non-uniform $\rho$}
\begin{figure}
\includegraphics[width=3.5in]{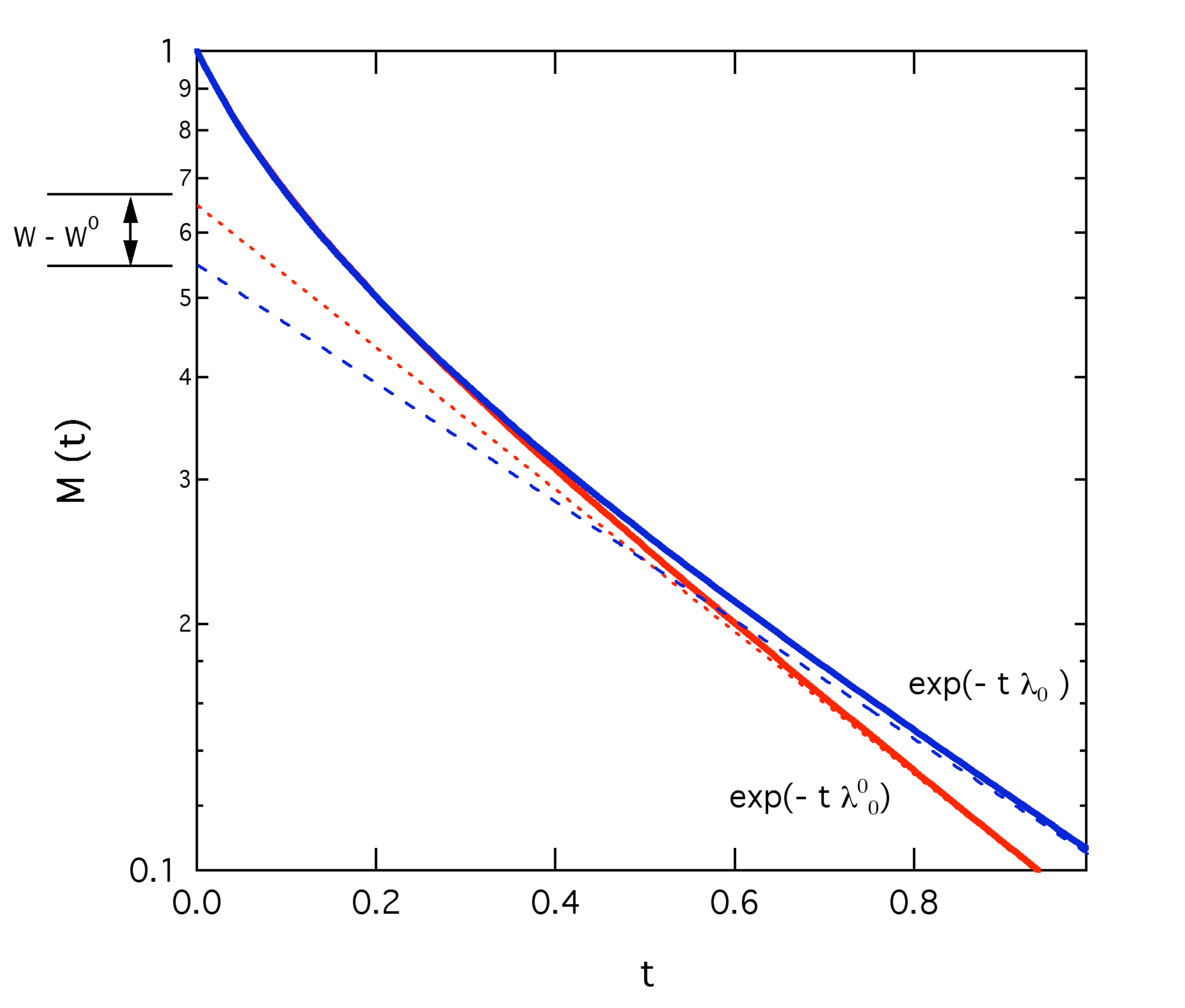}
\caption{\label{fig:wide}Schematics for the difference between the population evolution with uniform $\rho_0$ and an inhomogeneous $\rho(\r)$ with the finial slopes given by $\lambda_0^0$ and $\lambda_0$ as indicated by the broken curves. $W - W^0$ represents the change in the spectral weight distribution.}
\end{figure}

In this section, we derive expressions for the changes in the eigenvalues and their modes when the boundary condition varies from point to point. For this, we should introduce another set of eigenmodes $\{\phi_p\}$ with eigenvalues $\{\lambda_p\}$ as opposed to the superscripted eigensystem $\{ \phi_q^0 \}$ and $\lambda_q^0$'s for uniform $\rho(\r) = \rho_0$.
Using the self-adjointed property of ${\cal H}$, all eigenvalues are shown to be real, and their  associated eigenmodes may be represented as real functions, as we choose to do so in the following. 
Each $|\phi_p>$ now satisfies  
\begin{equation}
\label{eq:eigenequation}
{\cal H} | \phi_p > = \lambda_p | \phi_p > 
\end{equation}
and the non-uniform boundary condition
\begin{equation}
\label{eq:newbc}
<\r | \hat{n} (\r) \cdot {\bf J} | \phi_p>  =  < \r  |\rho(\r)| \phi_p>  \quad {\rm for}\, \r \in \Sigma.
\end{equation}
Following the steps that led to Eq.\ref{eq:theorem}, we obtain
\begin{equation}
\label{eq:theorem2}
\lambda_p = << \phi_p  | \rho (\r) |  \phi_p >> +  <\phi_p | {\bf J} \cdot D^{-1} \cdot  {\bf J} | \phi_p>.
\end{equation}
Our primary interest is now on the {\it difference} between the two eigensystems, without and with spatial variations in $\rho$,  
represented in terms of $\{ |\phi_q^0> \}$'s. 
Figure \ref{fig:wide} shows the schematics of changed properties in comparison to the uniform $\rho_0$ case, as we derive the details in this section.

With the definition of $\delta \rho$ and $\sigma$ given in Eq.\ref{eq:definedeltarho},  
We start by assuming that the eigenmode $| \phi_p>$ of the inhomogeneous case be expressed as a perturbation of the corresponding mode in the homogeneous counterpart, $|\phi_p^0>$: 
\begin{equation}
\label{eq:introdeltaphi}
|\phi_p>  = c_p (  | \phi_p^0> + | \delta \phi_p> ).
\end{equation}
One may further decompose   $| \delta \phi_p>$ into two components:
\begin{eqnarray}
\label{eq:phin}
|\phi_p > &=& 
c_p \Big(  |\phi_p^0> + \sum_{q\ne p} a_{pq} | \phi_q^0> \Big) + [I - {\cal P} ] |  \phi_p> 
\nonumber \\
&\equiv& c_p [ \sum_{q}^{all} a_{pq} | \phi_q^0> + | Q_p> ]
\end{eqnarray}
where ${\cal P}\equiv \sum_q^{all} | \phi_q^0> < \phi_q^0|$ is a projection operator into the Hilbert space spanned by the set of eigenmodes $\{ |\phi_q^0>\}$. $a_{pp} = 1$ by definition. 
The normalization condition $<\phi_p | \phi_p> = 1$ gives
\begin{equation}
\label{eq:normalization}
c_p =  (  \sum_{q}^{all} a_{pq}^2 + <Q_p | Q_p > )^{-1/2}.
\end{equation}
$|Q_p>$, which by definition should satisfy $<\phi_q^0 | Q_p> = 0$ for all q's, represents the part of $|\phi_p> $ that cannot be accounted for via a linear superposition of $|\phi_p^0>$'s. 
That $|Q_p>$ is not a null function, despite the fact that one can realize a least-square fit approximation\cite{Morse1953} with any given precision to an arbitrary function {\em inside} ${\cal V}$,  follows since the set $\{\phi_p^0 \}$ satisfies 
a boundary condition (Eq.\ref{eq:uniformbc}) while $\{ \phi_p \}$ satisfies another (Eq.\ref{eq:newbc}). 
Strictly speaking, any linear combination of $\phi_p^0$'s,  without the $|Q_p>$, cannot satisfy the {\em inhomogeneous} boundary condition, while at the same time, the role of $|Q_p>$ is probably secondary to $\sum_{q\ne p} a_{pq} |\phi_q^0>$, which needs to be verified.
Therefore, in the following, we first focus on getting $a_{qp}$'s  in terms of $\lambda_q^0$'s and the overlap integrals of $\delta \rho(\r)$ and $\phi_q^0$'s. 
We then evaluate $|Q_p>$ which satisfies
\begin{equation}
\label{eq:bcforqn}
 \frac{1}{c_p} \delta \rho(\r) |\phi_p> 
- \hat{n}  \cdot {\bf J} | Q_p> +  \rho_0 |Q_p>  = 0 
\end{equation}
on the boundary and is a solution to the {\em inhomogeneous} equation:
\begin{eqnarray}
\label{eq:eqforQn}
 ({\cal H} - \lambda_p) |Q_p>  &=& 
  \sum_{q} (  \lambda_p  -  \lambda_q^0 ) a_{pq} |\phi_q^0>  .
\end{eqnarray}
To be systematic, we obtain an approximate $|Q_{p,m}>$ with the perturbative solution $a_{p,q}$'s at the given stage ($\delta\rho^m$ with $m=1, 2, \ldots$ ) substituting for the {\em inhomogeneous source} term on the right hand side. It is assumed that $|Q_{p,m}> \rightarrow |Q_p>$ as $m \rightarrow \infty.$

Substitute Eq.\ref{eq:phin} into Eq.\ref{eq:eigenequation}, and multiply both sides by $<\phi_q^0|$, 
\begin{equation}
\label{eq:matrixeq}
\lambda_p^0 a_{pq} + <\phi_q^0 | {\cal H} | Q_p> = \lambda_p a_{pq}.
\end{equation}
From the self-adjointedness of ${\cal H}$ and the boundary condition, we make the crucial observation:
\begin{equation}
\label{eq:commutation}
<\phi_q^0 | {\cal H} | Q_p> =  c_p^{-1} << \phi_q^0 | \delta\rho | \phi_p >> 
\end{equation}
(See Appendix-\ref{appendix2}) where  
\begin{equation}
\label{eq:definesurfaceintegral}
<< \phi_q^0 | \delta\rho | \phi_p >> \equiv  \oint \phi_q^{0}(\r)  \delta \rho (\r) \phi_p (\r) d\sigma .
 \end{equation}
Using this, one can show 
\begin{equation}
\label{eq:matrixeq2}
(\lambda_p - \lambda_q^0) a_{pq}= (\sum_{r } a_{pr} \delta\rho_{qr} + \delta\tilde \rho_{qp})  \frac{\cal S}{{\cal V}}
\end{equation}
where we introduce the surface overlap integrals $\delta\rho_{qr}$ and $\delta \tilde \rho_{qp}^m$:
\begin{equation}
\label{eq:defmoment}
\delta\rho_{qr} \frac{\cal S}{{\cal V}} \equiv <<\phi_q^0 | \delta \rho | \phi_r^0> >
\end{equation}
\begin{equation}
\label{eq:defqmoment}
\delta \tilde \rho_{qp}^m \frac{\cal S}{{\cal V}}  \equiv << \phi_q^0 | \delta \rho| Q_{p,m}> >.
\end{equation}
It  formally follows that 
\begin{eqnarray}
\label{eq:gamma}
a_{pq} &=& \frac{1}{c_p}\frac{(1 - \delta_{pq})}{\lambda_p - \lambda_q^0} << \phi_q^0 | \delta \rho | \phi_p>>  + \delta_{pq} 
\end{eqnarray}
which, using Eq.\ref{eq:phin}, 
gives a recursive equation for $a_{pq}$:
\begin{eqnarray}
\label{eq:gamma1}
a_{pq }&=&   \frac{(1-\delta_{pq}) }{\lambda_p - \lambda_q^0} \Big\{ \sum_r a_{pr} \delta\rho_{qr} + \delta \tilde\rho_{qp}  \Big\}  \frac{\cal S}{{\cal V}} + \delta_{pq}.
\end{eqnarray}
Upon first iteration, and using the notation $\delta\tilde{\rho}^m_{qp}$ associated with the perturbative approximations $|Q_{p,m}>$, we obtain 
\begin{eqnarray}
\label{eq:gamma2}
a_{pq }&=&  \delta_{pq} +   (1-\delta_{pq}) 
 \Big\{ (
\frac{\delta\rho_{qp}  }{\lambda_p - \lambda_q^0}  
+\frac{ \delta \tilde\rho^1_{qp} }{\lambda_p - \lambda_q^0} )  \frac{\cal S}{{\cal V}} + \nonumber \\
&&(  \sum_{r\ne p}  \frac{ \delta\rho_{qr}}{\lambda_p - \lambda_q^0}   \frac{  \delta\rho_{rp} }{\lambda_p - \lambda_r^0}
+  \sum_{r\ne p} \frac{\delta\rho_{qr}   }{\lambda_p - \lambda_q^0}   \frac{\delta \tilde\rho^2_{rp} }{\lambda_p - \lambda_r^0}   + \nonumber \\
&&  \sum_{r\ne p}  \sum_{s \ne p} \frac{ \delta\rho_{qr}}{\lambda_p - \lambda_q^0}   \frac{  \delta\rho_{rs} }{\lambda_p - \lambda_r^0} a_{ps} ) ( \frac{\cal S}{{\cal V}} )^2
\Big\} 
\nonumber
\end{eqnarray}
providing a way for a systematic expansion in powers of $\frac{\delta \rho}{\rho_0}$ in a manner analogous to the diagrammatic expansion of a particle interacting with the perturbative potential. 
Truncating the iterations at second order,  we obtain, 
\begin{eqnarray}
\label{eq:gamma3}
a_{pq }&\approx&  \delta_{pq} +   (1-\delta_{pq}) 
 \Big\{ 
(\frac{\delta\rho_{qp}  }{\lambda_p - \lambda_q^0}  
+\frac{ \delta \tilde\rho^1_{qp} }{\lambda_p - \lambda_q^0} ) \frac{\cal S}{{\cal V}} + \nonumber \\
&& \sum_{r\ne p}   \frac{ \delta\rho_{qr}}{\lambda_p - \lambda_q^0}  
  \frac{  \delta\rho_{rp} }{\lambda_p - \lambda_r^0} ( \frac{\cal S}{{\cal V}} )^2
  + {\cal O}(\delta \rho^3)
\Big\} 
\end{eqnarray}
since $\delta \tilde\rho^m_{rp} \delta\rho_{qr} = {\cal O}(Q_{p,m}) \times \delta \rho^{2}$ or higher.
To obtain a formal solution in an algebraically closed form, we put $p=q$ in Eqs.\ref{eq:matrixeq2}  and obtain 
\begin{eqnarray}
\label{eq:eqforlambda}
\lambda_p - \lambda_p^0 
&=& \delta\rho_{pp}  \frac{\cal S}{{\cal V}} -  \sum_{q\ne p} \frac{\delta \rho_{pq}  \delta\rho_{qp}}
{ \lambda_q^0 - \lambda_p}  ( \frac{\cal S}{{\cal V}} )^2+  \delta \tilde\rho_{pp}  \frac{\cal S}{{\cal V}}  \nonumber \\
&+& 
 \sum_{q\ne p} \frac{\delta \rho_{pq}  << \phi_q^0 | \delta \rho | \delta \phi_p>>}{\lambda_p - \lambda_q^0} \frac{\cal S}{{\cal V}}  )
\end{eqnarray}
where 
$|\delta \phi_p> = \sum_{q\ne p} a_{pq} |\phi_q^0> + |Q_p>
$
and therefore the last term in Eq.\ref{eq:eqforlambda} contributes terms of order ${\cal O}(\delta\rho^3)$ and higher.
Keeping only up to second order in $\delta \rho$, and using the boundary condition for $|Q_p>$, we transform Eq.\ref{eq:eqforlambda} into the following alternative form: 
\begin{equation}
\label{eq:eqforlambda2}
\lambda_p - \lambda_p^0
= \delta\rho_{pp}  \frac{\cal S}{{\cal V}} -  \sum_{q\ne p} \frac{\delta \rho_{pq}  \delta\rho_{qp}}{ \lambda_q^0-\lambda_p} ( \frac{\cal S}{{\cal V}} )^2
 -\frac{1}{\rho_0}  ( S_{pp}  -  T_{pp} )  \frac{\cal S}{{\cal V}} 
\end{equation}
where we define 
\begin{equation}
\label{eq:definedeltarhosq}
S_{qp}  \frac{\cal S}{{\cal V}} \equiv \oint \phi_q^0 (\r) (\delta\rho(\r))^2 \phi_p^0(\r) d\sigma 
\end{equation}
and 
\begin{equation}
\label{eq:definet}
T_{qp}  \frac{\cal S}{{\cal V}} \equiv \oint \phi_q^0(\r) \delta \rho(\r) \hat{n}\cdot {\bf J} Q_p(\r) d\sigma .
\end{equation}
and take the $p = q$ case. 
Note that $S_{pp}$ assumes the form of a self-interaction which, through the mixed-boundary condition, is mitigated by the $T_{pp}$ term. We show in the following (section \ref{sec:sphere})  that, for the spherical pore geometry,  these two terms largely cancel each other. It suggests that the reduction in $\lambda_p$ gained through the self-interaction term ($S_{pp}$) is  lost due to the rapid bending of the mode profile ($T_{pp}$) near the boundary. 

We also have the generalization of the result we obtained for the spectral weight for the excited modes (Eq.\ref{eq:excitedfraction}):
\begin{eqnarray}
\label{eq:excitedfraction2}
W  &=&  
\Big( \sum_{q\ne 0} |s_q^0|^2 + {\cal V}  \sum_{q\ne 0} a^2_{0q} <   (\delta \phi_q^0)^2>  + \nonumber \\ && {\cal V} <  \delta Q_0^2>  - 2 {\cal V} ( <\phi_0^0>  \sum_{q\ne 0} a_{0q} <\phi_q^0>  +  \nonumber \\
&& <\phi_0^0><Q_0> + \sum _{q\ne 0} a_{0q} <\phi_q^0><Q_0>)  \Big)  \nonumber \\ 
&& / ( \sum_{q} a_{0q}^2 +   < Q_0 | Q_0 >  )
\end{eqnarray}
where we used the orthogonality properties of the $\phi_q^0$ and $Q_0$ and defined
$< (\delta A)^2 > \equiv \frac{1}{{\cal V}} \int_{{\cal V}}A^2 d\r -( \frac{1}{{\cal V}} \int_{{\cal V}} A d\r )^2$
and 
$<  A > \equiv \frac{1}{{\cal V}} \int_{{\cal V}} A  d\r $
where $A$ stands for $Q_0, \phi_q^0$ or $\phi_0^0$. 

To make further progress beyond Eq.\ref{eq:eqforlambda}, we need to solve for $|Q_p>$ and its surface integrals with respect to $\delta \rho(\r) \phi_q^0(\r)$.
Assuming that  $|\phi_p^0>$ and $\lambda_p^0$ are known for $\rho_0$, obtaining  $|Q_p>$, $a_{pq}$ and $\lambda_p$ in a mutually consistent manner  constitutes the complete solution of the problem which is not possible for a general pore geometry.  Instead, we are interested in 
how changes in observable properties such as the slowest eigenvalue and its spectral distribution depend on the texture of  $\delta \rho (\r)$ and the spatial profile of the relevant eigenmodes that reflect the underlying boundary geometry via the $\kappa$ parameter in a perturbative scheme based on small $\sigma$.

As noted with Eq.\ref{eq:gamma3} and Eq.\ref{eq:eqforlambda2}, if we restrict ourselves to the second-order perturbation evaluation of $\delta \lambda_{p}$, we need only to construct  $|Q_{p,1}>$ (i.e. $|Q_p>$ evaluated using the first order perturbative solutions)  as the  contribution of $Q_{p,m}$ to $\delta\lambda_p$ for $m=1,2,\ldots$ is $<<\phi_q^0| \delta \rho | Q_{p,m}>> \sim {\cal O}(\delta \rho^{m+1})$ or higher. We noted earlier that $|Q_{p,1}>$ arises from a distribution of source that is the remnant of $\delta \rho(\r)$ as $\sum_q a_{pq} |\phi_q^0>$ alone fails to account for its effect completely. 
Using Eq.\ref{eq:eqforQn} and Eq.\ref{eq:matrixeq2}, we obtain the following {\it inhomogeneous} Helmholtz equation for $Q_{p,1}(\r)$ to lowest order in $\sigma$:
 \begin{equation}
\label{eq:eqforQn3} 
({\cal H} - \lambda_p ) | Q_{p,1} >=   |f_{p,1}>
\end{equation}
where the source term is now given by 
\begin{equation}
\label{eq:source}
|f_{p,1} > \equiv  
 \frac{1}{c_p} \sum_{q}^{all}  | \phi_q^0><< \phi_q^0 | \delta \rho | \phi_p^0>>.
 \end{equation}
By extending the surface localized function $\delta \rho(\r)$ into a thin shell $(\partial V_\epsilon)$ of thickness $\epsilon$ lining the interface in the pore space, and ensuring that $\epsilon \ll \sqrt{D/\lambda_p^0}, \sqrt{D/\lambda_q^0}$,  we define
\begin{equation}
\label{eq:defineshell}
\delta \rho_\epsilon (\r) = 
\Big\{ 
   \begin{array} {ccc} 
      \epsilon \delta \rho (\r') & {\rm for } \, \r \in \partial V_\epsilon &  (\hat{\r'} \parallel (\r - \r'), \r'  \in \Sigma )  \\
      0 & {\rm otherwise} &  \\
   \end{array}.
 \end{equation}
to show that Eq.\ref{eq:source} is equivalent to
\begin{eqnarray}
\label{eq:source2}
| f_{p,1} > &=&  \lim_{\epsilon\rightarrow 0}
 \sum_{q}^{all}  | \phi_q^0> < \phi_q^0 | \delta \rho_\epsilon| \phi_p^0>
 \nonumber \\
 &=&  {\cal P}\delta \rho | \phi_p^0>, 
 \end{eqnarray}
the surface localized source function $\delta \rho(\r) \phi^0_p (\r)$ projected onto the space spanned by the eigenmodes $\{ |\phi_p^0> \}$.
In Appendix-\ref{appendix3}, we show that  $|Q_p>$  is given by the superposition of waves $G_p$ emanating from the {\em residual} charge $\sigma_{p,res}$:
\begin{equation}
\label{eq:compactstokesfinal}
|Q_p>  = (1 + G_p \delta \rho)^{-1} G_p  |\sigma_{p,res} >
 \end{equation}
where  $G_p$ is the Green's function as defined in Eq.\ref{eq:eqforgreenp}. 
For the second-order perturbation,  the residual charge is defined as
\begin{equation}
\label{eq:residual}
 |\sigma_{p,res} > =  {\cal P}\delta \rho |\phi_p^0> - \delta \rho {\cal P}  \frac{1}{c_p} |\phi_p>
\end{equation}
which is not necessarily limited to be on the interface. 
In the perturbative scheme, $|\phi_p>$ should now be replaced with $\sum_q a_{pq} |\phi_q^0> + |Q_{p,1}>$. 
The surface overlap integral of $|Q_p>$ and $\delta\tilde\rho_{qp}$ (Eq.\ref{eq:defqmoment}) can now be put into the following form:
\begin{eqnarray}
\label{eq:defqmoment2}
\delta\tilde\rho_{qp} \frac{{\cal S}}{{\cal V}}  &=&  \oint \oint d\sigma_1  d\sigma_3  \int d\r_2 \phi_q^0 (\r_1)  \delta \rho (\r_1) \times \nonumber  \\  
&& G_p (\r_1, \r_2)  \Big( {\cal P}(\r_2, \r_3) \delta\rho(\r_3) \phi^0_p(\r_3) -  \nonumber \\ 
&& \delta\rho(\r_3){\cal P}(\r_3, \r_2) \frac{1}{c_p} \phi_p(\r_2)  \Big)
 + {\cal O}(\delta \rho^3).
\end{eqnarray}
Closed solutions for $G_p$ and the residual charge for a general pore shape and $\delta \rho(\r)$ are not readily available, but 
evaluating $G_p$ out of the basis set $\{ \phi_q^0 \}$, and using orthogonality of $\{ \phi_q^0 \}$, we further obtain
\begin{equation}
\label{eq:solutionforQn3}
|Q_p> = - \sum_q \sum_{r\ne p} \frac{ |\phi_q^0> }{\lambda_q^0 - \lambda_p}  << \phi_q^0 | \delta \rho | \phi_r^0>>  a_{pr}
\end{equation}
which indicates that the requirement $<\phi_q^0|Q_p> \sim 0$ is obeyed in ${\cal O}(\delta \rho^2)$.
$Q_p(\r)$ may be interpreted as the potential field induced by the residual charge distribution $\sigma_{p,res}$ and its contribution to the eigenvalue $\lambda_p$ is the interaction between the potential and the surface charge distribution associated with each mode, $\delta \rho (\r) \phi_q^0(\r)$. The potentail is subject to a destructive interference when $\sigma_{p,res}(\r)$ has a rapid spatial fluctuation, and further weakened due to averaging over the diffusion length $\ell_p \sim \sqrt{\frac{D }{ \lambda^0_p}}$. (Eq.\ref{eq:eqforgreenp} in the appendix)
The leading order  contribution of $Q_p$ to $\lambda_p$, $\delta\tilde\rho_{pp}$ of Eq.\ref{eq:eqforlambda},  is then
\begin{equation}
\label{eq:qppfinalform}
\delta\tilde\rho_{pp}\frac{{\cal S}}{{\cal V}}= -  \sum_q \sum_{r\ne p} a_{pr}   << \phi_r^0 | \delta \rho | \phi_q^0>> \frac{<< \phi_q^0 | \delta \rho | \phi_p^0>> }{\lambda_q^0 - \lambda_p}  .
\end{equation}

Finally, we arrive at a compact expression 
\begin{equation}
\label{eq:fractionalshiftsmalldeltarho}
\delta \lambda_p = \delta\rho_{pp} \frac{{\cal S}}{{\cal V}}- \sum_{q\ne p} \frac{\delta\rho^2_{qp}}{\lambda_q^0 - \lambda_p } (\frac{{\cal S}}{{\cal V}})^2 + \oint \delta \rho(\r) \phi_p^0 (\r) Q_p(\r)  d\sigma
\end{equation}
for the change in eigenvalue valid up to second order in $\delta \rho$.  For the slowest mode, 
we conjecture that $|Q_{p=0}>$, as the surface source  $\delta \rho \, \phi^0_0$, is significantly weakened via convolution and commutation with ${\cal P}$ and the oscillatory kernel of wavelength $\ell_0$.
Taking $p=0$, we obtain the fractional shift in the decay rate of the slowest mode:
\begin{eqnarray}
\label{eq:fractionalshiftsmalldeltarho2}
\frac{\delta \lambda_0}{\lambda_0^0} &=& \frac{ \delta\rho_{00}}{\lambda_0^0} \frac{{\cal S}}{{\cal V}} -  
 \sum_{q\ne 0} \frac{\delta\rho_{0q}\delta\rho_{q0}}{\lambda_0^0(\lambda_q^0-\lambda_0^0)} (\frac{{\cal S}}{{\cal V}})^2   - \frac{1}{\lambda_0^0} \sum_q \sum_{r\ne 0}  \times \nonumber \\ 
 && \frac{\delta \rho_{0r}}{\lambda_r^0 - \lambda_0^0}  \delta \rho_{rq} \frac{\delta \rho_{q0}}{\lambda_q^0 - \lambda_0^0} (\frac{{\cal S}}{{\cal V}})^3 + \ldots
\end{eqnarray}
Note that the last term is of order ${\cal O}(\delta \rho^3)$ unless $q=0$. Therefore, taking only the $q=0$ contribution and rearranging, 
\begin{eqnarray}
\label{eq:fractionalshiftsmalldeltarho3}
\frac{\delta \lambda_0}{\lambda_0^0} &=& \frac{ \delta\rho_{00}}{\lambda_0^0} \frac{{\cal S}}{{\cal V}} -  
 \sum_{q\ne 0} \frac{\delta\rho_{0q}\delta\rho_{q0}}{\lambda_0^0(\lambda_q^0-\lambda_0^0)} (\frac{{\cal S}}{{\cal V}})^2 + \nonumber \\ 
 && \frac{1}{\lambda_0^0}  \frac{\delta \rho_{00}}{\delta \lambda_0}  \sum_{r\ne 0} \frac{\delta \rho_{0r}}{\lambda_r^0 - \lambda_0^0}  \delta \rho_{r0} (\frac{{\cal S}}{{\cal V}})^3 + {\cal O}(\delta \rho^3).
\end{eqnarray}
Due to the presence of $\delta \lambda_0$ in the denominator of the last term, $|Q_0>$ therefore makes a second order correction to the first term, only if $\delta \rho_{00} \ne 0$. Otherwise, its effect vanishes to second order in $\delta \rho$. 
We thus arrive at:
\begin{eqnarray}
\label{eq:fractionalshiftsmalldeltarho4}
\frac{\delta \lambda_0}{\lambda_0^0} &=& \frac{ \delta\rho_{00}}{\lambda_0^0}  \frac{{\cal S}}{{\cal V}} \Big( 1 +  \sum_{q\ne 0} \frac{\delta\rho_{0q}\delta\rho_{q0}}{\lambda_0^0(\lambda_q^0-\lambda_0^0)} (\frac{{\cal S}}{{\cal V}})^2 \Big)  -  \nonumber \\
&& \sum_{q\ne 0} \frac{\delta\rho_{0q}\delta\rho_{q0}}{\lambda_0^0(\lambda_q^0-\lambda_0^0)} (\frac{{\cal S}}{{\cal V}})^2 + {\cal O}(\delta \rho^3).
\end{eqnarray}

The fractional change in the weight for the excited mode (Eq.\ref{eq:excitedfraction2}) can now be simplified: 
\begin{eqnarray}
\label{eq:excitedfractionapprox}
\frac{W - W^0}{W^0}  \sim \frac{ \sum_{q\ne 0}(\frac{\delta\rho_{q0}  }{\lambda_0^0 - \lambda_q^0}  \frac{\cal S}{{\cal V}} )^2  <   (\delta \phi_q^0)^2> }{  <   (\delta \phi_0^0)^2>}
\end{eqnarray}
noting that  $<Q_0> \sim 0$, and $<\delta Q_0^2 > \sim < Q_0^2 > \propto {\cal O}(\delta\rho^4) $. We also expect that the fluctuations induced by $\delta \rho$ in the bulk of the pore when averaged over the pore volume will tend to vanish, $< \sum_{q} a_{0q} \phi_q^0>  \sim  0$, leaving the positive definite term above. 
(More precisely, one can show that these cancellations arise rigorously from the normalization condition, Eq.\ref{eq:trivialsumrule})
This shows that as the boundary condition becomes more inhomogeneous, the excited modes gains in weight in proportion to their mean-square fluctuation, $<   (\delta \phi_q^0)^2> $, but also weighted down by the $1/(\lambda_0-\lambda_q^0)^2$ factor and $\delta\rho_{q0}^2$, overlap between $\phi_q^0$ and $\delta \rho \phi_0^0$. The overall effect, however, is second order in $\delta \rho$ at most ($\sim a_{0q}^2$). 
These results are schematically summarized in Figure \ref{fig:wide}.

Following the steps taken for the uniform case, one can also show that the 
initial slope of population decay with a finite $\sigma$ satisfies:
\begin{equation}
\label{eq:initialslopeinhomogeneousrho}
\lim_{t\rightarrow 0} \frac{-1}{{\cal M}(t)} \frac{d}{dt} {\cal M}(t) = \rho_0 \frac{S}{{\cal V}} + \frac{\oint  \delta \rho(\r)  \Psi_t  (\r) d\sigma } {
\int  \Psi_t  (\r)  d\r }.
\end{equation}
As the second term vanishes for sufficiently short $t$ when $<\r | \Psi>_t$ is still uniform, it suggests that {\em the initial slope may remain robust}. Note however that the way the eventual deviation off the initial slope sets in may be different from that of the uniform $\rho_0$ case (even though the asymptotic value remains the same at $\rho_0 S / {\cal V}$) depending on how the depletion of population population proceeds.

We also note that the sum-rule that we found for the uniform case (Eq.\ref{eq:sumrule0}) retains the same form, 
and can be put into the form:
\begin{eqnarray}
\label{eq:sumrule2}
\sum_q^{all}   \frac{ |s_q^0|^2  |1 + \frac{ \delta s_q}{s_q^0} |^2 \lambda_q^0 (1 + \frac{ \delta \lambda_q}{\lambda_q^0}) } {1+ 2 \sum_{r \ne q} a_{rq}^2 + {\cal V}   < Q_q^2 >  } = \rho_0 \frac{S}{{\cal V}}
\end{eqnarray}
with $ \delta s_q = <\delta \phi_q | \Psi>_0 = \sum_{r\ne q} a_{qr} s_r^0 + <Q_q | \Psi>_0$.
Here $s_r^0$ is the overlap integral of $|\phi_r^0>$ with the initial distribution (defined earlier Eq.\ref{eq:spectraldistribution}) and $|\delta\phi_q>$ is as defined in Eq.\ref{eq:phin}.
Combining this with Eq.\ref{eq:sumrule0}, and taking the leading order in $\delta \rho$ only, 
we have the following condition that has to be satisfied
\begin{equation}
\label{eq:highordersumrule}
\sum_q^{all} \lambda_q^0 s_q^0 (\delta s_q + \frac{1}{2} s_q^0 \delta \rho_{qq}  )
 = 0.
\end{equation}
Note that it is in fact simply expressing the {\it rigidity} of the initial slope against the variational effect of  $\delta \rho$. 
To be complete, we note also that 
\begin{equation}
\label{eq:trivialsumrule}
\sum_q^{all} s_q^0 \delta s_q  =  0
\end{equation}
which  follows from the normalization requirement.
\section{Spherical Pore}
\label{sec:sphere}
As a solid example, let us consider the case of an isolated spherical pore. Although it is a three dimensional object, much of its MR response reduces to that of a one-dimensional system with a single controlling length scale. This is often overlooked and its properties have been casually interpreted as generic to pores with complex three dimensional morphology. To be more concrete, we sketch out analytic expressions and numerically evaluate them for some of the properties for the spherical pore despite this reservation. As an intermediate step, extension of the methods developed here to non-spherical pore geometry\cite{Finjord2007} would be useful. Angular variations in the boundary condition bring in aspects of the extra dimension and length scales more explicitly, (such as the terms contributing to Eq.\ref{eq:phin} with $L>0$) and it would be interesting to study the effect  on $\kappa$ (Eq.\ref{eq:generaldefinekappa}) of $\phi_0^0$ as it varies along the boundary of non-trivial geometry. Yet, to make an impact on $\delta \lambda_0 / \lambda_0^0$, their effects need to survive the angular-averaging; furthermore, the associated eigenmodes should have sizable presence on the boundary of the pore. 
Therefore, one should approach with discretion the conclusions we draw from the spherical pores in the following.

The eigenmodes for the uniform $\rho_0$ inside the spherical pore are separated into the radial part ($j_{L}( k  r)$) and the angular part ($Y_{LM}(\theta, \phi)$).  Instead of the generic index $q$, we  employ the set of indices $(n,L, M)$ that characterize the eigenmode associated with the eigenvalue $\lambda^0_{nL} \equiv D k_L(n)^2$ in terms of its radial and angular parts: 
\begin{equation}
\label{eq:sphereeigenmodes}
\phi^0_{n,L,M} (\r) =  c_{n,L} j_{L} (k_L(n) r) Y_{LM} (\Omega)
\end{equation}
where $j_L(k_L(n) r)$ is the spherical Bessel function associated with the angular momentum $L$. (Note that the modes defined thus are complex. However, the eigenvalues are all real, and the results so far remain valid.) $k_L(n)$ denotes the radial functions each associated with the $n-$th nodal, positive solution of
\begin{equation}
\rho_0 j_L(\alpha_L(n))  + D k_L(n)  j'_L (\alpha_L(n)) = 0
\end{equation}
where the dimensionless parameter $\alpha_n = k_L(n) a$ 
and $c_{n,L}$ is the normalization constant so that $\int_{\cal V} |\phi^0_{n,L,M}(\r)|^2 d\r = 1$. 
Since $\int d\Omega Y_{LM} Y^*_{L'M'} = \delta_{L,L'}\delta_{M,M'}$, it follows\cite{Abramowitz1972,Grebenkov2007a}
\begin{equation}
\label{eq:definecn}
c_{n,L}=\frac{\sqrt{2}\alpha_L(n)}{\sqrt{(\kappa - \frac{1}{2})^2 + \alpha_L(n)^2 - (L+\frac{1}{2})^2}} \frac{1}{j_L( \alpha_L(n) )} a^{-3/2}.
\end{equation}
In the analysis of Brownstein-Tarr\cite{Brownstein1979} and its subsequent application to a variety of porous media, all modes with $L\ne 0$ are excluded from consideration because all relevant integrals vanish under the uniform boundary condition and the isotropic initial state($L = 0, M=0$).  Here, we consider the non-uniform  $\delta \rho(\r)$ parametrized via its overall strength $\sigma$ and the angular variation on the sphere of radius $a$, $f(\Omega)$:
\begin{equation}
\label{eq:definefomega}
\delta \rho (\r)  \equiv \sigma \rho_0  f(\Omega)
\end{equation}
with which the modes with a finite angular momentum contribute to $\delta \lambda_0$  to further slow down the slowest mode.  Note that, in a similar manner, these higher eigenmodes play increasingly significant role as the pore geometry further deviates and acquires more asymmetry and heterogeneity.
Due to the $L=0$ symmetry of the mode $|\phi_{0}^0>$, the first order term in Eq.\ref{eq:fractionalshiftsmalldeltarho2} vanishes, and for the second order term, only states with the $L, M-$ components that are present in the $\delta(\r)$ profile contribute. 
Thus for the spherical pore, we have, up to second order in $\delta \rho$, 
\begin{equation}
\label{eq:sphereshift}
\frac{\delta \lambda_0}{\lambda_0^0}  = \frac{\delta \lambda_{0,a}}{\lambda_0^0} +  \frac{\delta \lambda_{0,b}}{\lambda_0^0} 
\end{equation}
where $\frac{\delta \lambda_{0,a}}{\lambda_0^0}$ is the contribution from coupling to the eigenmodes $\{ |\phi_{q}^0> \} (q \ne p)$ and $\frac{\delta \lambda_{0,b}}{\lambda_0^0}$ coming from $|Q_0>$.

\begin{figure}
\includegraphics[width = 3.5in]{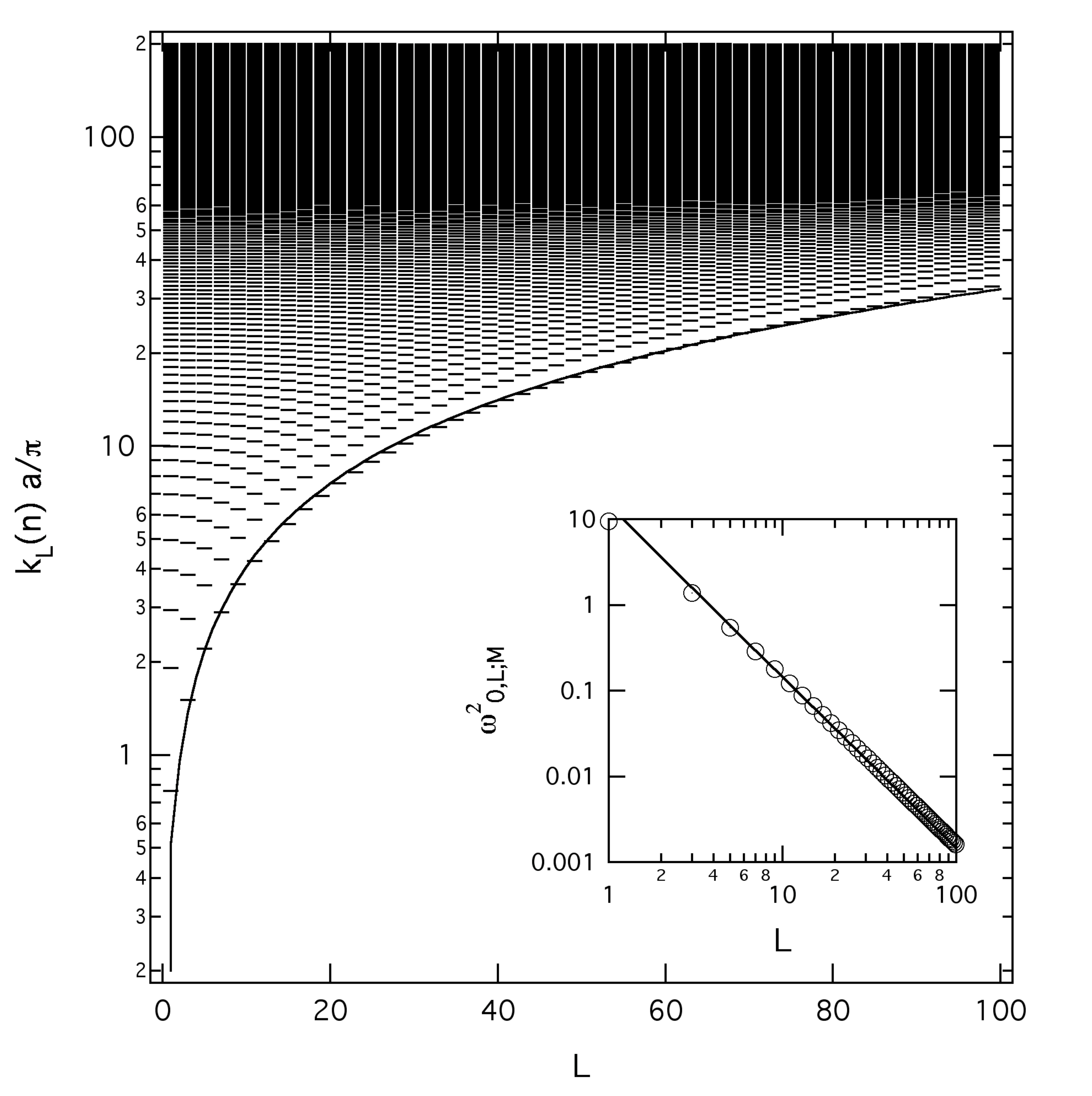}   
\caption{\label{fig:eigenvalues} 
An example of eigenvalues $k_L(n)$ for odd angular momenta $L =1, 3, \ldots, 99$ shown for $n$ up to 200 for $\kappa = 0.4161.$ The solid line ($k_L(1) a / \pi \sim  0.51 L^{0.9}$) is just a guide for the eyes. Shown in the inset is the angular factor $\omega^2_{0,L;M}$ (Eq.\ref{eq:defineomega}) for the hemi-spherical $\rho(\r)$ variation.}
\end{figure}

\begin{figure}
\includegraphics[width = 3.2in]{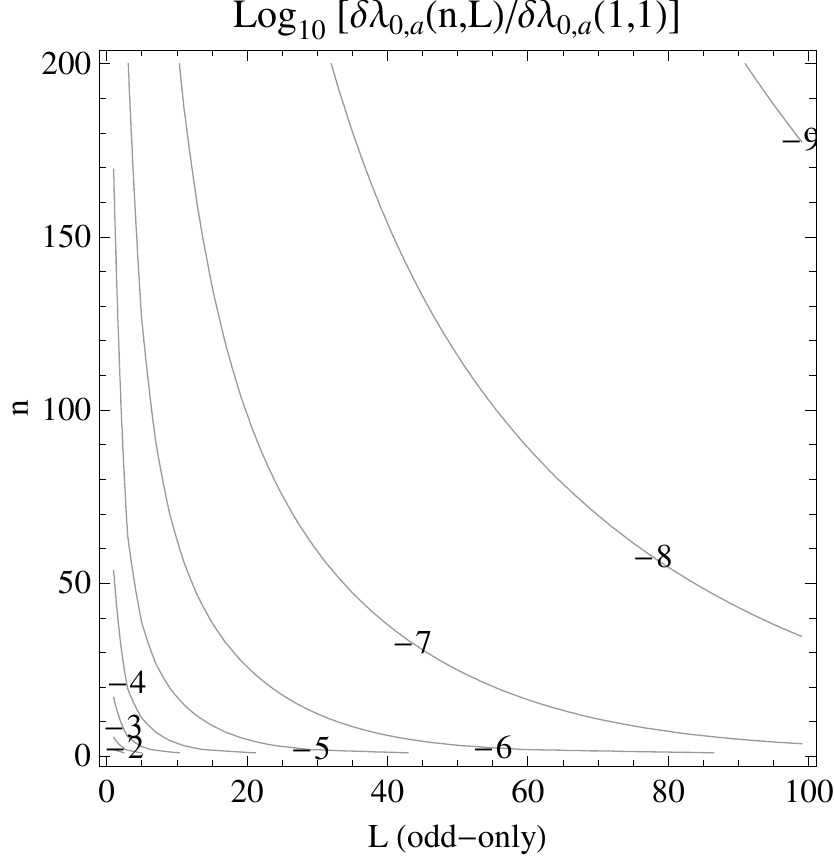} 
\caption{\label{fig:deltalambda} 
Contribution to $\delta \lambda_{0,a}$ from each eigenmode with radial node index $n = 1,2,\ldots$ and angular momenum $L (=1,3,5,\ldots)$ for $\kappa= 0.416.$ The contour levels represent base-10 logarithm of individual contribution $\delta \lambda_{0,a}(n,L)$ normalized to the maximum value $\delta \lambda_{0,a}(1, 1) = 0.0323.$
}
\end{figure}

\begin{figure}
\includegraphics[width = 3.5in]{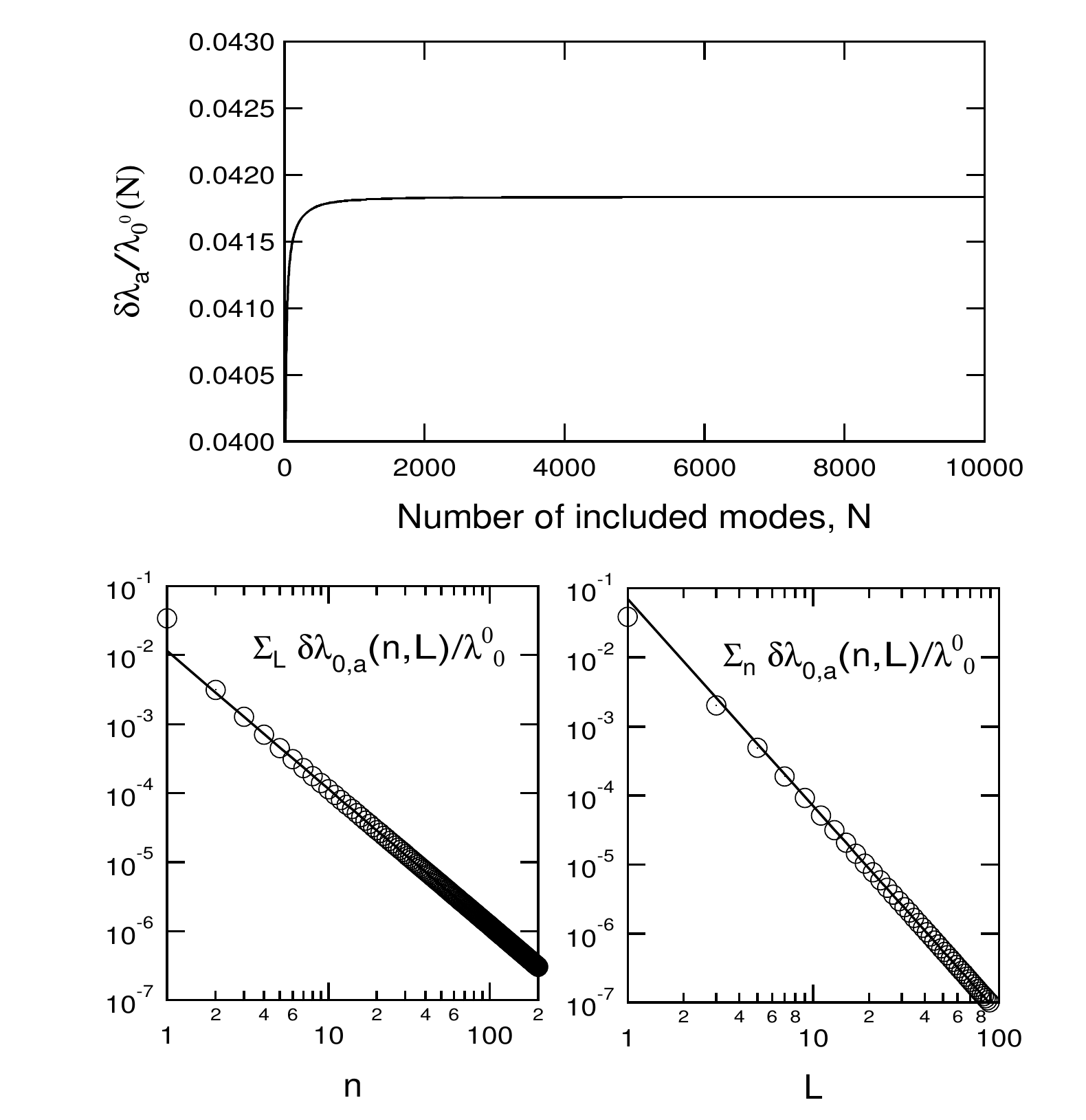} 
\caption{\label{fig:deltalambdaGraphs} 
Upper panel: Convergence of numerically evaluated $\delta \lambda_{0,a}$ as total number of modes increases. All eigenmodes in the ranges $1 \le n \le 200$ and $1 \le L \le 101$ were found, and their individual contribution $\delta \lambda_{0,a}(n, L)$ evaluated and sorted according to their magnitude. The graph shows $\sum_{i}^N \delta \lambda_{0,a} (i)$ as the number of included modes $N$ increases. 
Lower panels: Dependence of partially summed $\delta \lambda (n,L)$ on $n (=1,2,\ldots)$ and $L$ for $\kappa= 0.416. $The left panel shows contribution from all modes with same $n$ summed over $L \le 101$.  
The right panel shows contribution with same $L$ values summed over all $n \le 200$. The solid lines are guides for the eyes with $\propto n^{-2}$ and $\propto L^{-3}$ respectively.
}
\end{figure}

Let us examine the two second order contributions one by one. 
The first term, $\frac{\delta \lambda_{0,a}}{\lambda_0^0}$, is
\begin{eqnarray}
\label{eq:fractionalshiftsmalldeltarho2sphere}
\frac{\delta \lambda_{0,a}}{\lambda_0^0} &\sim& -  
 \sum_{L,M,n}  c^2_{1,0} c^2_{n,L}  j_0(k_0(1) a)^2 j_{L}(k_L(n) a)^2 \times \nonumber \\
 &&  \frac{(\sigma \rho_0)^2 \omega_{0,L;M}^2}{\lambda_0^0(\lambda_{n,L}^0-\lambda_0^0)}\frac{a^4}{4\pi}.
\end{eqnarray}
where we define 
\begin{equation}
\label{eq:defineomega}
\omega_{0,L;M} \equiv   \oint d\Omega f(\Omega)Y_{LM}^{*}(\Omega) 
\end{equation}
as the $Y_{L,M}$ component in the harmonic expansion of $\delta \rho$.
Introducing the slowest rate for $\kappa \rightarrow \infty$, 
\begin{equation}
\label{eq:definelambdainfty}
\lambda_\infty = D (\frac{\pi}{ a})^2
\end{equation}
and using Eq.\ref{eq:definecn} and $\alpha_L(n) \equiv  k_L(n) a$, we can put this into a form which displays its dependence on $\kappa$ explicitly for an arbitrary angular variation of $\delta \rho$:
\begin{eqnarray}
\label{eq:fractionalshiftsmalldeltarho3sphere}
&& \frac{\delta \lambda_{0,a}}{\lambda_0^0} \sim -  
 \sigma^2 \kappa^2 \frac{1 }{\pi^5 } \sum_{L,M,n}  \omega_{0,L;M}^2
 \frac{\lambda_\infty}{\lambda_0^0}  \frac{\lambda_\infty}{(\lambda_{n,L}^0-\lambda_0^0)} \times \nonumber \\
 &&   \frac{\alpha_{L}^2(n)}{(\kappa - \frac{1}{2})^2 + \alpha_{L}^2(n) - (L+\frac{1}{2})^2}
 \frac{\alpha_{0}^2(1)}{\kappa (\kappa -1) + \alpha_{0}^2(1) }.
\end{eqnarray}
Let us consider a simple case with  
\begin{equation}
\label{eq:hemirhodefine}
\delta \rho (\r) \equiv \sigma \rho_0 f(\Omega) =  \sigma \rho_0
\Big\{ \begin{matrix} 
      -1 & (\theta \le \pi/2) \\
      1 & (\theta > \pi/2) \\
   \end{matrix}
\end{equation}
in which only $L=$ odd modes with $M=0$ are present. Figure \ref{fig:eigenvalues} shows the eigenvalues found for odd $L$ up to $10^2$, with $n \le 200$ for each $L$. Figure \ref{fig:deltalambda} shows how much individual eigenmode with $L,n$ contributes to $\delta\lambda_0$ while Figure \ref{fig:deltalambdaGraphs} shows the rapid convergence properties as the number of included mode increases. It also shows the partial contributions from all modes with a given $L$ or $n$ values. 

\begin{figure}
\includegraphics[width = 3.5in]{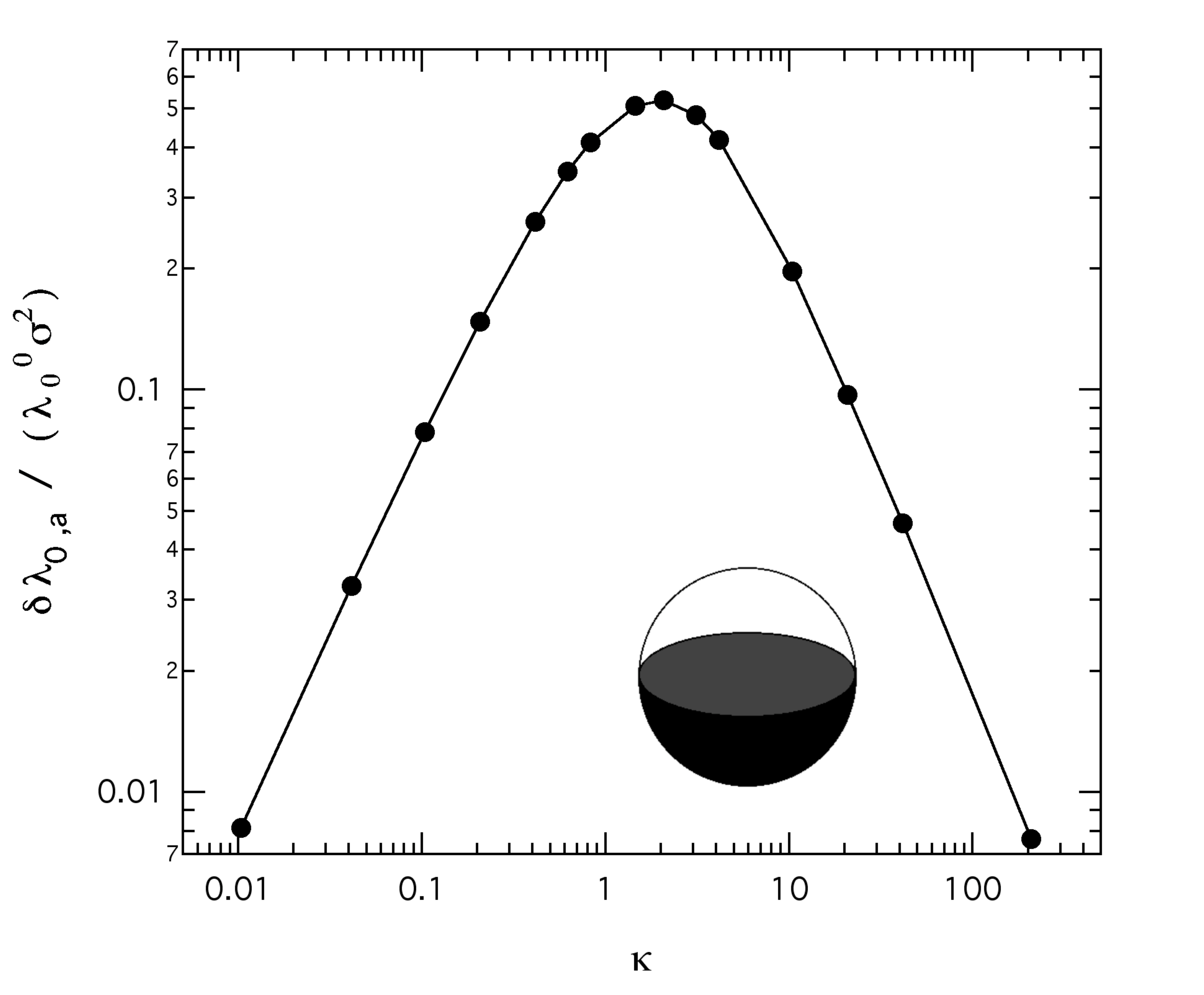}
\caption{\label{fig:dela} 
Second-order contribution to the fractional shift of the slowest relaxation rate ($\delta \lambda_{0,a}/\lambda_{0,0}$) for the hemispherical, binary distribution of $\rho(\r)$ for the spherical pore at various values of $\kappa$. Up to 20000 modes were included to ensure good convergence for all $\kappa$ values. The contribution peaks around $\kappa  \sim  2.$
}
\end{figure}
For small $\kappa \ll 1$, $\lambda_0^0/\lambda_\infty \propto \kappa$ and 
$\lambda_\infty / (\lambda_{n,L}^0 - \lambda_0^0)$ becomes independent of $\rho$, and therefore one can see $\delta\lambda_{0,a}/ \lambda_0^0 \sim \kappa$, while for $\kappa \rightarrow \infty$, we note that  
$\lambda_0^0/\lambda_\infty$ becomes independent of $\kappa$, therefore 
\begin{equation}
\label{eq:sumproperty1}
\sum_n \frac{\lambda_\infty}{\lambda_{n,L}^0-\lambda_0^0}  \frac{\alpha_{L}(n)^2}{(\kappa - 0.5)^2 + \alpha_{L}(n)^2 - (L+0.5)^2} \propto \frac{1}{\kappa}
\end{equation}
so that  $\delta \lambda_{0,a}^0/\lambda_0^0 \propto \kappa^{-1}$.
Figure \ref{fig:dela} and Table \ref{t:numbers} show the numerically evaluated $\delta\lambda_{0,a}/ \lambda_0^0$ for a wide range of $\kappa$ with the hemispherical $\delta \rho$ which bears out this observation. For a qualitative description, this may be roughly described as 
\begin{equation}
\label{eq:delavskappa}
\delta \lambda_{0,a}^0/\lambda_0^0 \sim \frac{2.3}{\kappa + \frac{4}{\kappa}} \sigma^2.
\end{equation}
The result suggests that, at least for the spherical pore, the system is most sensitive to the inhomogeneity in the intermediate range $\kappa \in [0.5, 10]$ in its second order contribution. 
Incidentally, this is where the two terms in the rate (as we noted earlier Eq.\ref{eq:theorem} and also in \cite{Ryu2001}) are comparable and the system becomes most accommodating of the perturbation $\delta  \rho$. If the former dominates (i.e. $\kappa \ll 1$), it becomes too costly for the slowest mode to deform itself from quasi-uniformity to accommodate $\delta \rho$, while in the opposite case, the mode amplitude near the boundary  is severely reduced (i.e.  $j_0(k_0(1) a),  j_L(k_L(n) a) \rightarrow 0$), and $\delta \lambda_0$ becomes insensitive to a fractional change in $\delta \rho$.

More generally, for an eigenmode  of angular variation with $(L,M)$ to contribute at least a fraction $\beta$ of $\delta \lambda / \lambda_0^0$, 
the strength of the corresponding component in the harmonic variation of  $\delta \rho$, $\omega_{0,L;M}$ would have to meet
\begin{equation}
c_{1,0}^2 c_{n,L}^2  j_0^2(k_0(1) a) j_{L}^2(k_L(n) a)\frac{(\sigma \rho_0)^2 \omega_{0,L;M}^2 }{\lambda_0^0(\lambda_{n,L}^0-\lambda_0^0)}  \frac{a^4}{4 \pi}>  \beta 
\end{equation}
which can be used to define the region of relevance in the $\{L, n\}$-plane for numerical evaluations. 
Even for pores without spherical symmetry, this criterion may be generalized using the strength of the associated modes averaged over the interface that should replace $c_{n,L}^2 j_L^2$ and $c_{1,0}^2 j_0^2$. It should be emphasized however, that, if the first order contribution survives,  this second order effect may become overshadowed.

\begin{table}[tb]
   \centering
\begin{ruledtabular}
   \begin{tabular}{c c c c c c} 
      $\kappa$  & $\frac{k_0(1) a}{\pi}$ & $c_{1,0}$ & $j_0(k_0(1) a)$ & $\frac{\lambda_0^0}{\lambda_\infty}$ & $ \frac{\delta \lambda_a}{\lambda_0^0} \frac{1}{\sigma^2}$  \\
\hline
 0.0104      & 0.056176 & 0.013868  & 0.99482 & 0.003158 & 0.008125\\
 0.0416      & 0.112001 & 0.013997  & 0.97949 & 0.012544 & 0.03250 \\
0.1040	& 0.17599 & 0.0142543 & 0.949825 & 0.030973 & 0.078438 \\
 0.2081      & 0.246325 & 0.014680  & 0.90314 & 0.060516 & 0.14750\\
 0.4161      & 0.341266 & 0.015517  & 0.81914 & 0.116281 & 0.26125\\
 0.6242      & 0.409545 & 0.016331  & 0.74606 & 0.16810 & 0.34750\\
 0.8323     & 0.463496 & 0.017118  & 0.68225 & 0.214369 & 0.41107\\
 1.4565     & 0.57835 & 0.01929  & 0.53379 & 0.334489 & 0.50800 \\
 2.0807     & 0.65411 & 0.0211704  & 0.430702 & 0.42786 & 0.52330 \\
3.1210      & 0.736232 & 0.0236819 & 0.318651& 0.542038 & 0.48001\\
4.1614      & 0.788449 & 0.025566  & 0.24899 & 0.620944 & 0.417812\\
 10.403      & 0.906404 & 0.030683  & 0.10178 & 0.820836 & 0.196594\\
20.8069	& 0.95228	 & 0.0329654 & 0.0499212 & 0.90684 & 0.096940\\
41.6138	&0.976014&	0.034195& 0.02455 & 0.952603 & 0.0465482 \\
 208.07     & 0.995194 & 0.035206  & 0.004829 & 0.990411 & 0.0076395 \\
  \end{tabular}
\end{ruledtabular}
   \caption{Numeric values for $k_0$, $c_{1,0}$, $j_0(k_0 a)$, $\frac{\lambda_0^0}{\lambda_\infty}$ and $\frac{\delta\lambda_a}{\lambda_0^0}$ for $\kappa = 0.01$ to $200$ obtained using summation up to 20000 eigenmodes.}
   \label{t:numbers}
\end{table}

Evaluation of the second term of Eq.\ref{eq:sphereshift}, $ \frac{\delta \lambda_{0,b}}{\lambda_0^0},$ is more involved for an arbitrary pore geometry, as we need to evaluate first the ${\cal P}$ projection of the surface localized function $\delta \rho(\r)$ and further its overlap integral with $G_{p=0}(\r_1, \r_2)$, neither of each is available in a closed form. 
For spheres, however, we can clearly see from Eq.\ref{eq:fractionalshiftsmalldeltarho3}  that it should make a vanishing contribution as 
$<< \phi_0^0 | \delta \rho | \phi_0^0>> = 0$ as $<\r|\phi_0^0> \propto Y_{00}(\Omega)$. 
Comparisons to an exact solution\cite{Johnson2008} and numerical simulations\cite{Ryu2008d} verify  that 
\begin{equation}
\label{eq:dlambdabvanishes}
\frac{\delta\lambda_{0,b}}{\lambda_{0}^0} = 0 \quad {\rm if } \, \oint d\sigma \phi_{0}^0 \delta \rho(\r) \phi_0^0(\r) = 0
\end{equation} 
for the second order contribution of $Q_0$ to $\delta \lambda_{0}$ in spherical pores with an arbitrary $\delta \rho(\r)$.
At the same time, it is plausible that  there exist  boundary shapes for which $\phi_0^0(\r)$ develops a significant angular variation so that $\oint \phi_0^0 \delta \rho \phi_0^0 d\sigma \ne 0$. In such a case, the first order contribution in Eq.\ref{eq:fractionalshiftsmalldeltarho4}  would dominate. $|Q_p>$ only contributes to its higher order modification. 
It is instructive to examine how the projection ${\cal P}$ and the residual source $|\sigma_{0,res}>$ behave in more detail. 

\begin{figure}
\includegraphics[width = 3.5in, height = 2.6in]{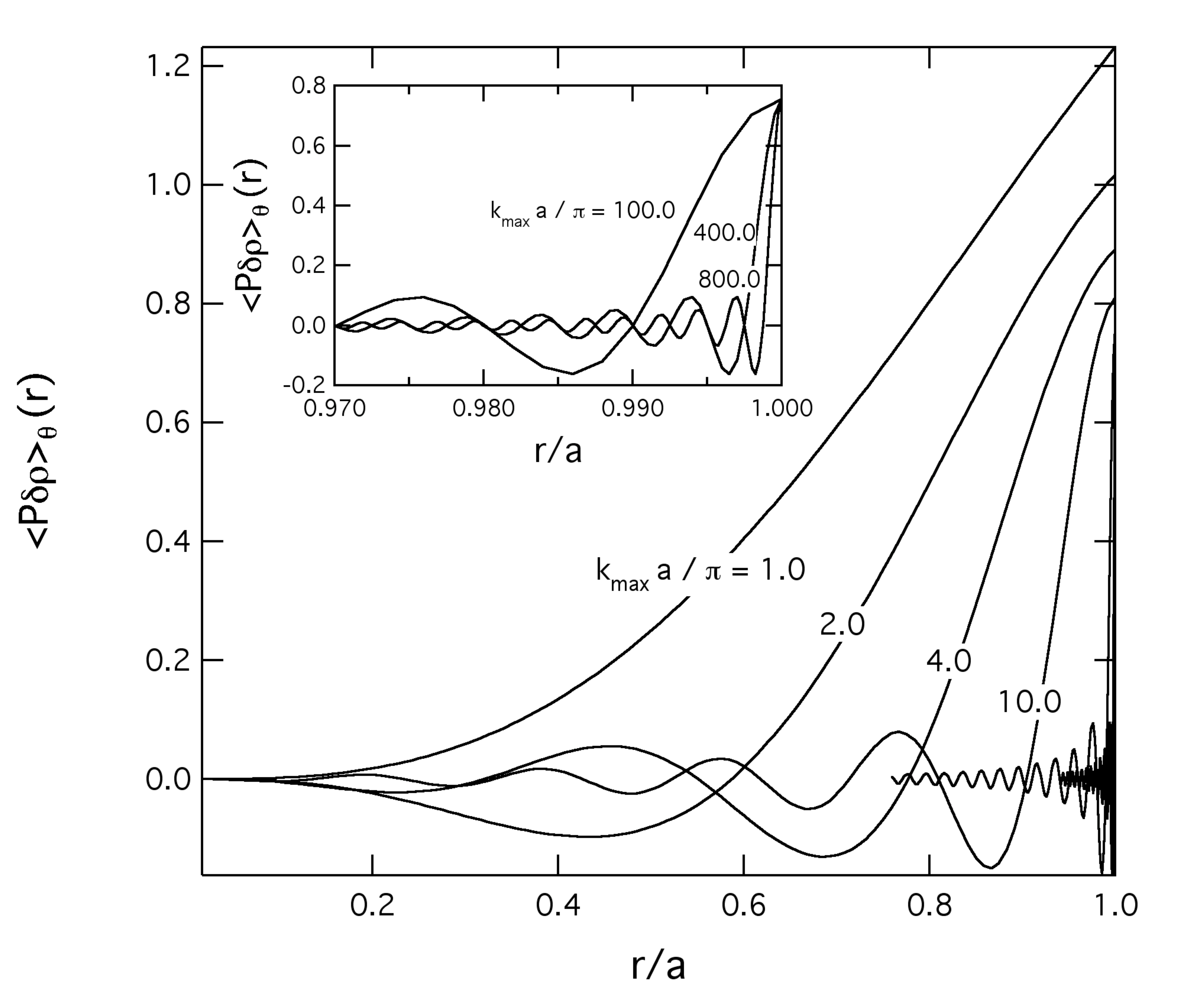}
\caption{\label{fig:deltaradial} 
Radial profile of the {\it partially-}projected, $\theta-$averaged  $<{\cal P}_{L,n_m} \delta \rho>_\theta (r)$ (the curves are multiplied by $(\frac{r}{a})^2 \frac{1}{4} \frac{2 \pi}{k_{max}a}$ for normalization) near the pore edge where $L=1$ only and $n_m = k_m a / \pi$ as indicated for each curve up to $800$ (see inset for blow-up for large $n_{m}$).  The height of the normalized curves at $r/a = 1$ converges toward $0.75075$ for $\kappa = 0.4$ in this example.  With inclusion of more modes with $L > 1$, the height approaches the value of $1.0$ gradually as one would expect for a perfect representation. The convergence becomes progressively slower as $\kappa$ increases. (See Fig.\ref{fig:deltafunctions}).
}
\end{figure}

Let us first consider the numerical evaluation of the projected $\delta \rho$, ${\cal P} f(\Omega)$ of the binary distribution (Eq.\ref{eq:hemirhodefine}).  
For a delta-profile for $\delta \rho(\r)$ in the radial direction with $\frac{\partial}{\partial r} \delta \rho(\r) = 0$ at the boundary, ${\cal P} \delta \rho$, as evaluated numerically with a large number of $\phi_q^0$'s, may approach $\delta \rho$ with an arbitrary precision\cite{Morse1953}, and yet fail to meet the zero-slope condition since all $\{ \phi_q^0 \}$'s have a finite slope on the boundary unless $\rho_0 = 0$. This discrepancy may hardly impact the accuracy of the surface integrals $\oint d\sigma \phi_q^0 \delta \rho(\r) \phi_0^0$ in practice. 
Discrepancy between $\delta \rho$ and its projection ${\cal P} \delta \rho$ may become pronounced along the $(d_e-2)$ dimensional manifold (i.e. the equatorial line $\theta = \pi/2$ in the hemispherical example) across which sharp changes in $\delta \rho$ occur at length scales smaller than $\ell_p$. How this translates into an enhanced contribution to $\delta \lambda_{0,b}$ can only be addressed numerically for a general $\delta \rho(\r)$ texture. In the following, we investigate how the numeric ${\cal P} \delta \rho(\r)$ representation behaves as a series-sum over a finite number of modes for the simple spherical model. 

Figure \ref{fig:deltaradial} shows how the radial delta-function like profile is approached with progressively larger number of radial modes (with cutoffs $k_{max}$ as indicated) averaged over $\theta \in [0, \pi/2]$. The oscillatiotory tail is due to the finite cutoff . The inset shows details near the boundary for larger cutoff values. 
With only $L=1$ modes included, the value on the boundary converges to the value of $\sim 0.75$, significantly short of $1.0$. Inclusion of $L>1$ modes remedies this, but its convergence is significantly impeded as $\kappa$ increases. 

\begin{figure}
\includegraphics[width = 3.3in, height = 2.4in]{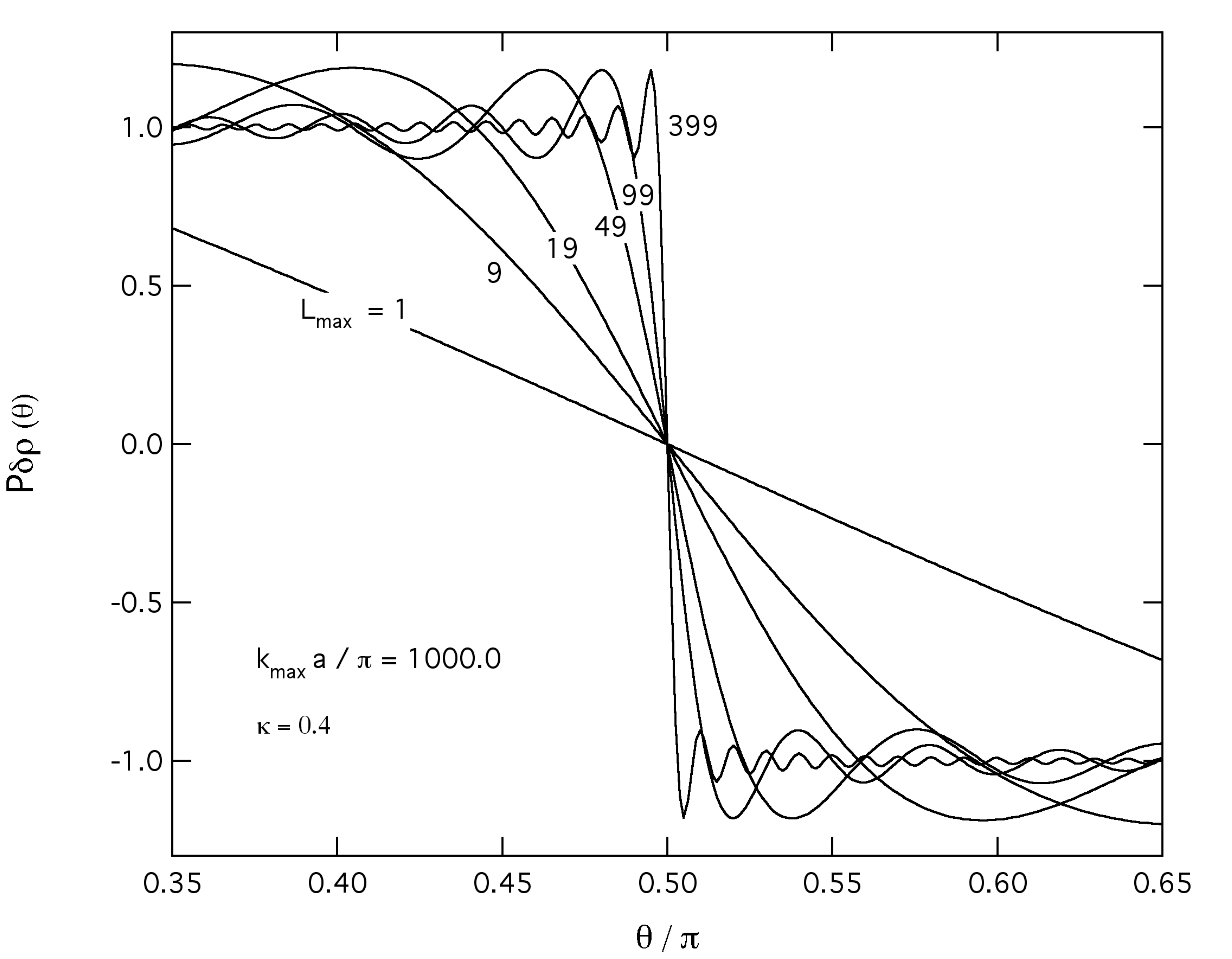}
\caption{\label{fig:deltatheta} 
Strength of the projected ${\cal P}_{L,n_m} \delta \rho$ around $\theta = \pi/2$ at $r=a$ across which it should go from $-1.0$ to $1.0$ in a stepwise fashion. The curves show progressive refinement as we increase the number of included modes by increasing the maximum $L$ up to $399$ ( $k_{L}(n_m) a / \pi$ was fixed at $1000$ for each included $L$).
}
\end{figure}
Figure \ref{fig:deltatheta} shows how the angular profile (Eq.\ref{eq:hemirhodefine}) is reproduced with progressively larger number of angular modes (with cut-off $L_{max}$ as indicated). The radial cutoff is set at $k_{max} a / \pi = 1000$.
In the hemispherical case, a moderate value of $L_{max}$ seems sufficient to achieve an acceptable convergence, although we observe that its rate slows down as $\kappa$ increases. 

\begin{figure}
\includegraphics[width = 3.5in]{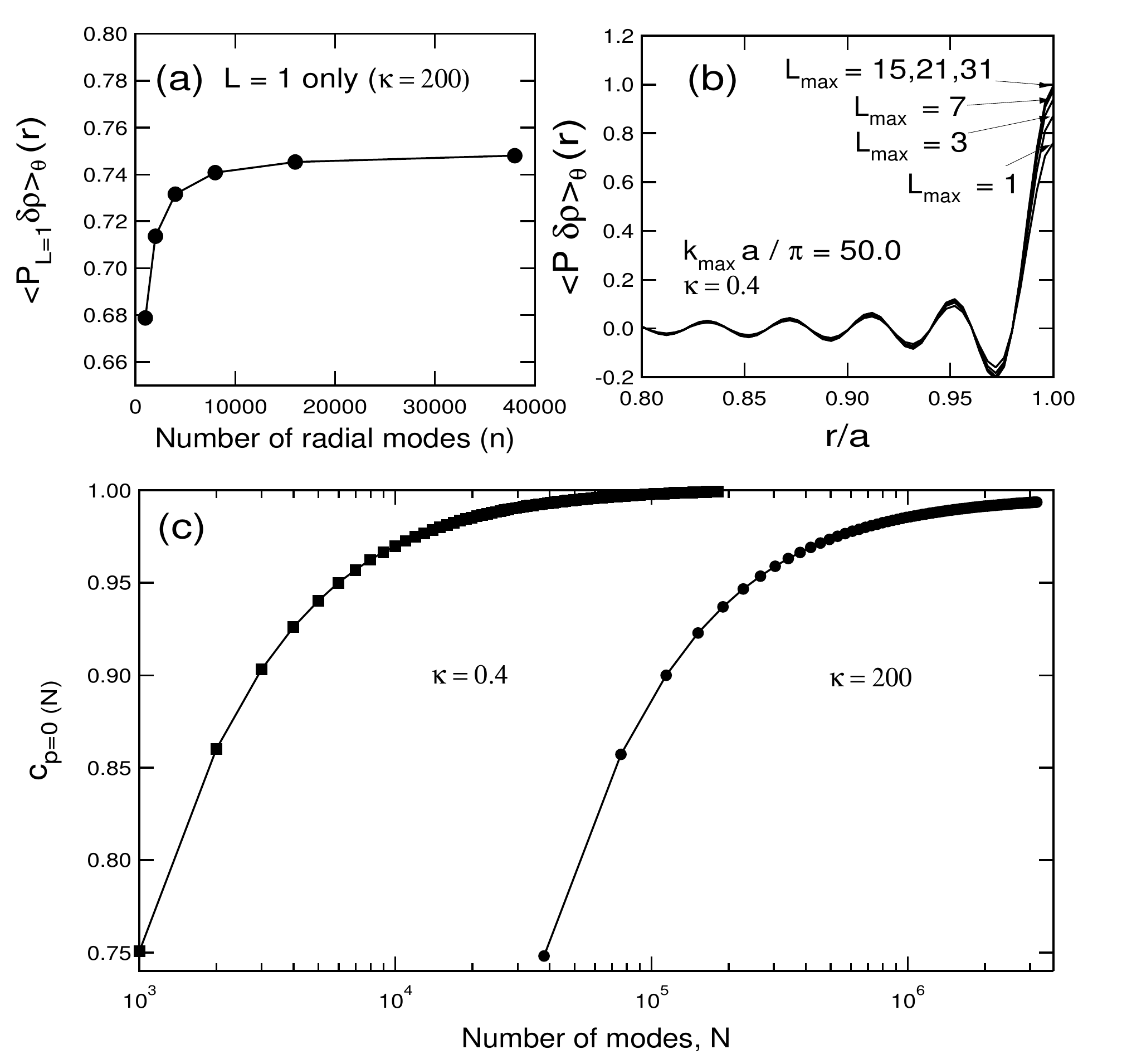}
\caption{\label{fig:deltafunctions} 
Convergence of the projected $\delta \rho \phi_0^0$ to the actual. 
Panel (a) shows that the value of the partial projection ($L=1$ modes only) with $\kappa = 200$, 
averaged over the hemispherical shell approaches $\sim 0.746$, falling short of the expected $1.0$ even after summing over 40000 radial eigenmodes. 
Panel (b), which shows the radial profile of the projected $\delta \rho \phi_0^0$ for $\kappa  = 0.4$, 
indicates that good convergence is achieved  when modes with $L$ up to $15$ or more are included. 
Panel (c) shows how the convergence $c_{p=0} \rightarrow 1.0$ slows down as $\kappa$ increases. With $\kappa = 200$, even after including up to $3\times 10^6$ modes, the convergence is not quite complete.
}
\end{figure}

We monitored the following dimensionless parameter as a measure of convergence for the ${\cal P}$-projected $\delta \rho$ to the actual $\delta \rho$ (as an overlap integral with $\phi_p(\r)$ over the pore-volume) in contributing to $\delta \lambda_{p,b}$:
\begin{equation}
\label{eq:definecpevaluationgeneral}
c_{p} \equiv \frac{
\int d\r_2 \sum_{q}  \phi_q^0(\r_2)  \oint d\sigma_3  \Theta( \delta \rho (\r_3) ) \phi_q^0 (\r_3) \delta\rho(\r_3)  \,   \phi_{p}^0 (\r_3)
}{ \oint d\sigma_2 \Theta( \delta \rho (\r_2))  \delta\rho(\r_2) \phi_{p}^0(\r_2)}, 
\end{equation}
where $\Theta(x)$ is the heavy side step function, $=1$ for $x>0$ and $=0$ otherwise.
Its convergence is largely determined by the spectral weight of modes $\{ \phi_{q\ne p}^0(\r) \}$ present in $\delta(\r)\phi_p^0(\r)$. 
To be systematic, we define the partial projection operator $
{\cal P}_{L,n_{m}} \equiv \sum_{n}^{n_{m}} |\phi_{n,L,0}><\phi_{n,L,0}|$ 
that projects onto a subspace spanned by the first $n_m$ radial modes for each $L$.
Panels (a) and (b) of Figure \ref{fig:deltafunctions} show the angular average of the 
partial projection $\sum_L^{L_{max} }{\cal P}_{L,n_{m}}\delta \rho | \phi_p^0>$ for values of $L_{max}$ and $n_m$ (equivalent to $k_{max}$). 
In panel (b), small but rapid oscillations observed in radial direction are immaterial as they are averaged out when convoluted over the pore volume. As the convergence of $c_p$ is largely controlled by the spectral compostion of the source profile, $f(\Omega)$, this may no longer hold for a complex $\delta\rho$ texture.\cite{note2}

For the hemi-spherical $\delta\rho$, $c_p$ for $p=0$ becomes
\begin{equation}
\label{eq:definecpevaluation2}
c_{p=0} =\lim_{n_{m}\rightarrow \infty}
\sum_{L} \epsilon_m \frac{\oint_{\theta < \pi/2}  d\sigma_2  \oint d\sigma_3 <\r_2| {\cal P}_{L,n_m}| \r_3>}
{\oint_{\theta < \pi/2} d\sigma_2}
\end{equation} 
where $\epsilon_m$ is the radial width of the projected delta peak on the boundary for the given $n_m$, 
and is given by the quarter of the wavelength associated with the mode with $k_L(n_m)$: $\epsilon_m = \frac{1}{4}\frac{2 \pi}{k_{L}(n_m)}$.
Panel (c) of Figure \ref{fig:deltafunctions} shows this $c_{p=0}$ as we progressively increase the number of modes in its numerical evaluation for both large and small values of $\kappa$. 
It is worth noting that while the convergence of  $c_p$ (and therefore $\delta \lambda_{0,b}$) is quite slow, the series sum for $\delta \lambda_{0,a}$ is much more rapid due to the factor of $\frac{1}{\lambda_q^0 - \lambda_0^0}$ (in Figure \ref{fig:deltalambda}). In a numerical simulation that  employs random walkers with a fixed step size, one would be effectively truncating the series summation at a wavelength comparable to the stepsize. The apparent strength of $\sigma$ in such simulations may then deviate from what corresponds to a fully converged $c_p$ in the figure. Our result provides a guide on how one may correct for such artifacts in a systematic manner. Numerical simulations employing large number of random walkers, which uses continuous step sizes to alleviate such an issue, are underway for various types of pores. 

\section{Conclusion}
We considered the consequence of the spatially varying boundary condition $[D(\r) \hat{n} \cdot \nabla + \rho(\r)] \Psi(\r,t) = 0$ for the  spatio-temporal evolution of the local density $\Psi(\r,t)$ of an attribute carried by diffusing entities. It has direct relevance on the local magnetic (polarization) density of fluid molecules in the magnetic resonance relaxometry widely used for various porous media for their characterization. We examined the spectral properties of the governing Helmholtz equation and their relationship to the boundary geometry and the texture of its controlling parameter $\rho(\r)$. 
Using only the general properties of the modes and the boundary conditions they satisfy, we showed that each eigenvalue can be expressed as a sum of two parts (Eq.\ref{eq:theorem}): one is in the form of a surface integral directly involving $\rho(\r)$, the other being a volume integral which involves the diffusive flux of the mode. 
The direct relationship between the slowest eigenvalue and the surface-to-volume ratio of the pore is recovered when the first term dominates over the second, and we derived the generalized parameter $\kappa$ (Eq.\ref{eq:generaldefinekappa}) that quantifies the boundary between distinct regimes with observable consequences in the evolution of an initial distribution.  We also showed that the weight of the slowest decay mode in the overall relaxation of the attribute is diminished in direct proportion to the rms spatial fluctuation of the mode (Eq.\ref{eq:excitedfraction}). Traditionally, the weight for all modes other than the slowest is often interpreted as representing {\it small pores}, a notion increasingly invalid as the pores become extended through diffusive coupling and acquire complex geometry. We clarified issues regarding the time-domain evolution of $\Psi(\r,t)$ which originate from such inadequate interpretation of the spectral distribution, $s_q^0$ (Eq.\ref{eq:spectraldistribution}). 
Building on this, we then introduced spatially varying  $\rho(\r)$ and obtained the perturbative solution in $\sigma = <|\delta \rho|> / \rho_0$. The results show how the effect of $\delta \rho(\r)$ manifests itself as the shift in the lowest eigenvalue and is controlled by overlap integrals of $\delta \rho(\r)$ with the associated mode. (Eq.\ref{eq:definesurfaceintegral}) We also showed and verified numerically that the initial slope of the overall depletion remains robust (See Eq.\ref{eq:highordersumrule}).
These results were derived without making any specific assumption about the pore geometry, relying only on the self-adjointedness and boundary conditions of the problem.  We show that the first order contribution vanishes when the base system has symmetry so that  the overlap integral with the mode $\phi_p^0$,  $\oint \phi_p^0(\r) \delta \rho(\r)  \phi_p^0(\r) d\sigma = 0$.  When the boundary geometry varies in a complex manner, the slowest mode $\phi_0^0(\r)$ itself acquires significant spatial variaion even under the uniform $\rho_0$, and the incommensuracy between $\delta \rho(\r)$ and $\phi_0^0$'s may result in a significant {\em non-zero} first order contribution (Eq.\ref{eq:fractionalshiftsmalldeltarho}), which overshadows the second-order effect.  
In the opposite limit, where the texture of $\delta \rho(\r)$ is such that its variation occurs on length scales much shorter than the pore geometrical length-scale $\ell$ (Eq.\ref{eq:defineell}), the effect of such {\em finely inhomogeneous} $\delta \rho$ should be muted via diffusive averaging-out. 
This is the case considered in Valfouskaya {\em et al} \cite{Valfouskaya2006} which considered a texture with a random variation {\em uncorrelated beyond the voxel size}, much smaller than the typical grain size, in a stochastically generated 3D porous medium. Our numerical simulations on random glass bead packs with similar textures of $\delta \rho$ also yielded results consistent with these observations.\cite{Ryu2008b} Towards the opposite limit, we applied out theory to a case where the impact of a finite $\sigma$ may be most pronounced (section \ref{sec:sphere}). Extensive numerical analysis including up to $3\times 10^6$ eigenmodes  was performed for the simple case of a spherical pore for a wide range of $\kappa = \rho_0 a /  D$. We propose that the fractional change in the slowest eigenvalue is the most effective probe into variations in the boundary condition. We examined how much each of the eigenmodes contributes to the change in the slowest rate (Figure \ref{fig:deltalambda}),  and obtained the overall second order contribution for a wide range of $\kappa$ (Fig \ref{fig:dela}).The latter is observed to peak around the value of $\kappa \sim 2.0$, and follows roughly $\propto \kappa$ and $\propto 1/\kappa$ at either end. Our result provides a useful theoretical framework and quantitative bounds for more complex situations addressed mainly through numerical simulations.\cite{Arns2006, Valfouskaya2006, Ryu2008b} Further development through comparison to exact solution\cite{Johnson2008} and systematic numerical simulations\cite{Ryu2008d} are underway. 

\begin{acknowledgments}
The author would like to thank Dave Johnson for helpful criticism and suggestions.
\end{acknowledgments}

\appendix

\section{}
\label{appendix2}
\noindent Here we show that 
\begin{equation}
<\phi_q^0 | {\cal H} | \delta \phi_p> - < \delta \phi_p | {\cal H} |\phi_q^0 > =  c_p^{-1} << \phi_q^0 | \delta\rho | \phi_p >> 
\end{equation}
where $|\phi_q^0>$ is an eigenmode with the uniform boundary condition, $|\phi_p>$ is an eigenmode with the inhomogeneous boundary condition, $c_p$, its normalization constant with $|\phi_p> = c_p (|\phi_p^0> +  |\delta \phi_p> )$ and ${\cal H} = \nabla \cdot {\bf J} $ with ${\bf J} = - D \cdot  \nabla$. 
For notational brevity, we choose the representation in which all the eigenmodes are real functions.
We start by putting the  first term on left hand side as
\begin{equation}
 \int_{{\cal V}}  \Big(  \nabla \cdot  \phi_q^0 (\r)  {\bf J}  \delta \phi_p (\r) - ({\bf J} \delta \phi_p (\r)) \cdot \nabla \phi_q^0(\r) 
 \Big)  d\r 
\end{equation}
and the second term as
\begin{equation}
\int_{{\cal V}} \Big(  \nabla \cdot  \delta \phi_p (\r )  {\bf J} \phi_q^0 (\r)
- ( {\bf J}  \phi_q^0(\r)) \cdot \nabla  \delta \phi_p ( \r) \Big)  d\r 
\end{equation}
and using the Gauss's theorem to put the left hand side into:
\begin{equation}
\label{eq:appendixtheorem}
\oint_\Sigma \Big(  \phi_q^0 (\r) \hat{n}\cdot {\bf J}  \delta \phi_p (\r)  - \delta \phi_p ( \r)  \hat{n}\cdot  {\bf J}  \phi_q^0 (\r) \Big)  d\sigma .
\end{equation}
Substituting $\delta \phi_p (\r) = c_p^{-1}  \phi_p (\r)- \phi_p^0 (\r)$, the integrand becomes 
\begin{equation*}
\phi_q^0 ( \r )  (  \hat{n}\cdot  {\bf J}  \frac{\phi_p (\r)}{c_p} - \hat{n}\cdot  {\bf J} \phi_p^0 (\r))  
-  ( \frac{\phi_p(\r)}{c_p}- \phi_p^0 ( \r)) \hat{n}\cdot  {\bf J}  \phi_q^0 (\r) 
\end{equation*}
which upon using the boundary conditions turns Eq.\ref{eq:appendixtheorem} into:
\begin{equation}
c_p^{-1}  \oint_\Sigma  \phi_q^0 ( \r )   \delta \rho(\r)  \phi_p (\r)  d\sigma .
\end{equation}
   
\section{}
\label{appendix3}
\noindent 
Here we derive the particular and homogeneous solutions for $|Q_p>$ at a given stage in the perturbative iteration and show how $Q_p$ is related to $\delta \rho$ and $|\phi_p^0>$.    
We start by noting that $Q_p (\r)$ is the solution to the inhomogeneous Helmholtz equation 
\begin{equation}
\label{eq:eqforqp0}
({\cal H} - \lambda_p) | Q_p > = | f_p >
\end{equation}
for the $|Q_p>$ function where $\lambda_p$ is the perturbative eigenvalue consistent with the boundary condition as satisfied by the solution from the previous stage,  $f_p(\r)$ is given accordingly in Eq.\ref{eq:eqforQn}. The boundary condition that should be satisfied by $Q_p$ is 
\begin{equation}
\label{eq:appendixbcforqn}
 \frac{1}{c_p} \delta \rho(\r) \sum_q a_{pq} \phi_q^0(\r)
- \hat{n} (\r) \cdot {\bf J} Q_p (\r) +  \rho(\r)   Q_p (\r)  = 0.
\end{equation}
Using the projection operator ${\cal P}= \sum_q | \phi_q^0> < \phi_q^0|$ onto the Hilbert space spanned by $\{ |\phi_q^0>$, $|Q_p>$ is related to $|\phi_p>$ via 
\begin{equation}
\label{eq:relationbetweenqpandphip}
|Q_p> = \frac{1}{c_p} (I - {\cal P}) | \phi_p>
\end{equation}
from which it follows that 
${\cal P}|Q_p > = 0$. 
Our aim is to seek a formal solution for $Q_p(\r)$ using the Green's function approach. 
Consider the following Green's function 
\begin{equation}
\label{eq:eqforgreenp}
({\cal H} -  \lambda_p) G_p(\r, \r_1) =\delta (\r - \r_1).
\end{equation}
We consider its representation in the basis functions of an eigen-system  $\{ |\xi_q> \}$ that satisfies
\begin{equation}
({\cal H} - \epsilon_{q}) |\xi_{q}> = 0
\end{equation}
and the general condition
\begin{equation}
(\rho_0  - \hat{n}(\r)\cdot  {\bf J} ) \xi_q (\r) =  \zeta(\r) \xi_q(\r)
\end{equation}
on the boundary with the function $\zeta(\r)$ to be chosen for convenience.  Choice of $\zeta (\r) = -  \delta\rho(\r) $ amounts to solving for $\{ \phi_q \}$,  while the lowest order perturbation would amount to the choice of $\zeta(\r) = 0$.

The Green's function  $G_p$ may now be given in terms of $|\xi_{q}>$:
\begin{equation}
\label{eq:greensfunction}
G_p(\r, \r') = < \r | \sum_q |\xi_q > \frac{1}{ \epsilon_q - \lambda_p^0} <\xi_q |  \r' >
\end{equation}
and it satisfies the boundary condition 
\begin{equation}
\label{eq:bcforgp}
(\rho_0 - \hat{n}(\r)\cdot  {\bf J}) G_p(\r, \r_1) =   \zeta(\r) G_p(\r, \r_1).
\end{equation}
To obtain $|Q_p>$ as a perturbative solution for the source function $f_p$ and $G_p$, 
we multiply Eq.\ref{eq:eqforqp0} by $G_p(\r_1, \r)$ from the left side and integrate over $\r$ to get
\begin{equation}
G_p {\cal H} | Q_p > - \lambda_p G_p | Q_p >  =  G_p | f_p>.
\end{equation}
Using Stoke's theorem, this becomes
\begin{eqnarray}
\label{eq:compactstokes2}
( {\cal H} G_p) | Q_p > &+&   G_p |_\epsilon {\bf J} | Q_p >   -  ( {\bf J} G) |_\epsilon Q_p>  - \lambda_p G_p | Q_p >\nonumber \\
  &=&  G_p | f_p>
\end{eqnarray}
where $|_\epsilon$ indicates a surface integral.
Replacing $( {\cal H} G_p) = \lambda_p G_p +I $ and using the boundary conditions for $Q_p$ (Eq.\ref{eq:appendixbcforqn})
and $G_p$ (Eq.\ref{eq:bcforgp}), after rearranging, we get 
\begin{equation}
\label{eq:compactstokes3}
\Big( I- G_p |_\epsilon (\delta \rho(\r) + \zeta(\r)) \Big) | Q_p >   = G_p | \sigma_{p,res} >
\end{equation}
where we introduce the {\em residual} source: 
\begin{equation}
\label{eq:appendixresidualsource}
|\sigma_{p,res}> \equiv | f_p > -  \sum_q a_{pq} \delta\rho  | \phi_q^0>.
\end{equation}

The general solution for $|Q_p>$ admits possible addition of the solution, $|Q_p^h>$, for the homogeneous counterpart of Eq.\ref{eq:eqforQn}. We now show that $|Q_p^h> = 0$ on physical grounds. 
First, note that the homogeneous equation for $|Q_p>$ and the original problem for $|\phi_p>$ becomes identical. The same applies to the boundary condition. This suggests we may take the homogeneous solution $|Q_p^h>$ to be identical to $|\phi_p>$ up to a constant factor $\alpha$,  $|Q_p^h>  = \alpha |\phi_p>$, with $\alpha$ to be determined by consistency requirements:
\begin{equation}
\label{eq:consistencyforQp}
\alpha ( {\cal P} |\phi_p> + |Q_p^p>  + |Q_p^h>) = |Q_p^h>
\end{equation}
and the boundary condition 
\begin{equation}
\label{eq:consistencyforQp2}
[-\hat{n}\cdot {\bf J} + \rho_0] (|Q_p^p> + |Q_p^h>) + \delta \rho \, \phi_p = 0
\end{equation}
where $|Q_p^p> $ is the particular solution of Eq.\ref{eq:compactstokes3}.
Since $|Q_p^p>$  term should vanish by its construction, and replacing $|Q_p^h>$ with $\alpha |\phi_p>$ on the left hand side turns the boundary condition into 
\begin{equation}
\label{eq:consistencyforQp3}
(1 + \alpha) \delta \rho (\r) \,  \phi_p (\r) = \delta \rho (\r) \, \phi_p (\r)
\end{equation}
which therefore requires $\alpha = 0$, i.e. we should take $|Q_p^h> = 0$.


\bibliography{strass}

\begin{thebibliography}{57}
\bibitem[{not()}]{note2}
\bibinfo{note}{As the angular variation of $\delta \rho$ becomes more rapid, it
  acts as a high-pass filter in controlling the strength of each mode with $L$
  to $\delta\lambda_{0,b}$ (as indicated by the lower bounding curve of Figure
  \ref{fig:eigenvalues}). In the case of the hemi-spherical $\delta\rho$, all
  odd angular momenta bigger than $L= 1$ contributes, and one would expect that
  the impact of this filtering would amount to suppression of some modes with
  $L\le1 $ and $n \sim 1$. For a distribution with rapid angular variations,
  the number of such suppressed modes may increase, thereby impacting the
  convergence of the projection.}
\end{thebibliography}

\end{document}